\documentclass[journal]{IEEEtran}
\IEEEoverridecommandlockouts
\usepackage{graphicx}
\usepackage[cmex10]{amsmath}
\usepackage{algorithmicx}
\usepackage{amsmath}
\usepackage{algpseudocode}
\usepackage{algorithm}
\usepackage{multirow}
\usepackage{color}
\usepackage{amsfonts}
\usepackage{verbatim}
\usepackage{cite}
\graphicspath{{fig}}
\usepackage{subcaption}



\newlength{\mywidth}
\makeatletter
\newif\ifhbonecolumn
\@ifclasswith{IEEEtran}{onecolumn}{\hbonecolumntrue}{\hbonecolumnfalse}
\makeatother
\ifhbonecolumn
\usepackage[top=1.0in, bottom=1.0in, left=1.0in, right=1.0in]{geometry}
\fi

\begin{document}
\title{
Multi-Agent Reinforcement Learning for Multi-Cell Spectrum and Power Allocation}
\author{Yiming Zhang, Dongning Guo
\thanks{The authors are with Department of Electrical and Computer Engineering, Northwestern University, Evanston, IL 60208, USA (e-mail: \{yimingzhang2026, dguo\}@northwestern.edu).}
\thanks{The work was supported in part by the National Science Foundation (NSF) under grant Nos.~2003098 and~2216970, a gift from Intel Corporation, and also the SpectrumX Center under NSF grant No.~2132700.}
}


\maketitle
\begin{abstract}
This paper introduces a novel approach to radio resource allocation in multi-cell wireless networks using a fully scalable multi-agent reinforcement learning (MARL) framework. 
A distributed method is developed where agents control individual cells and determine spectrum and power allocation based on limited local information, yet achieve quality of service (QoS) performance comparable to centralized methods using global information. The objective is to minimize packet delays across devices under stochastic arrivals and applies to both conflict graph abstractions and cellular network configurations. This is formulated as a distributed learning problem, implementing a multi-agent proximal policy optimization (MAPPO) algorithm with recurrent neural networks and queueing dynamics. This traffic-driven MARL-based solution enables decentralized training and execution, ensuring scalability to large networks.
Extensive simulations demonstrate that the proposed methods achieve comparable QoS performance to genie-aided centralized algorithms with significantly less execution time. The trained policies also exhibit scalability and robustness across various network sizes and traffic conditions.
\end{abstract}

\begin{IEEEkeywords}
Markov decision process;
multi-agent reinforcement learning (MARL);
recurrent neural networks;
stochastic traffic;
wireless networks.
\end{IEEEkeywords}

\section{Introduction}
\label{sec:Intro}

The increasing density of devices and access points (APs) in cellular networks, driven by growing consumer demands, has heightened the significance of coordinated resource allocation between cells. In this context, the AP in each cell needs to make multifaceted decisions, including which mobile device to serve in the downlink, at what time, using which sub-bands, and at what power levels. Our goal is to develop scalable, traffic-driven and {\em fully distributed} methods that achieve comparable quality of service (QoS) as those of well-known {\em centralized} methods, including the weighted minimum mean-squared error (WMMSE)~\cite{shi2011iteratively} and fractional programming (FP)~\cite{shen2018fractional}. By {\em fully distributed}, we refer to a system where each AP executes an algorithm that requires input from APs in at most a small neighborhood, so that a typical cell has fixed computational complexity even if the number of cells keeps increasing as the network expands.

To better understand the challenges in resource allocation, we first abstract the wireless communication network as a conflict graph, which effectively represents interference and constraints between links. In a conflict graph, centralized method like the Max Weight algorithm achieves optimal throughput~\cite{srikant2013communication}, but require identifying all maximum independent sets within the graph, which is an NP-complete problem~\cite{tarjan1977finding}. While Greedy Maximal Scheduling (GMS) provides a simpler alternative, it remains centralized and thus impractical for large networks. Low-complexity heuristic methods such as Longest-Queue-First (LQF) often support only a portion of the capacity region. As a more practical solution, the queue-length-based carrier-sense multiple access (Q-CSMA)~\cite{ni2011q} was proposed, offering improved performance over LQF while utilizing only local information.

We then consider a more practical cellular network model with analog channel states, where the resource allocation extends beyond scheduling to include power control for interference mitigation. However, existing approaches face various limitations. Centralized methods like WMMSE and FP require global CSI across the entire network, and their computational complexities scale rapidly with the network size. Heuristic methods such as random/full-power allocation, require minimal information but often support only a small portion of the capacity region, as they ignore inter-cell and intra-cell interference. ITLinQ~\cite{naderializadeh2014itlinq}, a low-complexity scheduling method, attempts to balance performance and simplicity by scheduling transmissions in subsets of links with ``sufficiently'' low interference levels. However, it still requires global CSI and coordination, as links sequentially decide whether to participate in scheduling. Distributed optimization approaches like~\cite{huang2006distributed,kiani2007maximizing} aim to avoid the extensive information exchange required by centralized methods, but often exhibit inferior performance compared to centralized methods due to partial or imperfect CSI.

The aforementioned centralized methods and distributed methods presents a clear trade-off between computational complexity, information exchange requirements, and performance. This balance leads to our motivation again: Is it possible to develop resource allocation methods that achieve QoS performance comparable to centralized approaches while only utilizing local information for decision-making? Such methods would be scalable and practical for large network deployments, aligning more closely with real industry needs.

Machine learning has recently emerged as a powerful tool for wireless resource allocation problems, offering potential solutions to this challenge. Supervised learning approaches, as demonstrated in~\cite{sun2018learning}, have trained deep neural networks using WMMSE-generated datasets to approximate its policy. In~\cite{zhao2021distributed}, graph neural networks (GNNs) have been adopted to leverage topological information for user scheduling in conflict graphs. 
Reinforcement learning (RL) offers advantages in avoiding high-dimensional, non-convex optimization, providing a model-free approach, and aligning well with sequential decision-making. Pioneering works applying deep RL to power control~\cite{nasir2019multi} achieved sum-rate performance closely matching that of FP and WMMSE algorithms. Further advancements have expanded RL applications to joint sub-band selection and transmit power control using deep Q-networks~\cite{tan2020deep} and actor-critic networks~\cite{nasir2021deep}. Multi-agent RL (MARL) introduces multiple agents that interact and learn simultaneously to achieve desirable rewards, with applications in power allocation~\cite{khan2020centralized} and MISO systems~\cite{ge2023deep}.


However, practical networks often face constraints on communication overhead or excessive delays, necessitating distributed approaches where agents make decisions based on limited local information. The aforementioned learning methods fall short in allowing truly distributed deployment with limited observations at each access point. Supervised learning methods~\cite{sun2018learning} learns from WMMSE, which requires global CSI. GNN approach~\cite{zhao2021distributed} requires global topology information. The methods in~\cite{nasir2019multi, tan2020deep, nasir2021deep} requires extensive CSI exchange between links. In~\cite{tan2020deep}, the use of Cartesian product action spaces also face convergence issues as sub-bands increase. In~\cite{khan2020centralized} and~\cite{ge2023deep}, authors make assumptions about independent transition functions and the reward is shared by all agents. The centralized training and distributed execution (CTDE) framework, common in these RL-based works, limits their scalability. While~\cite{chang2023federated} incorporates federated learning with MARL to enable distributed training, it still requires centralized parameter reporting and achieves lower performance compared to centralized methods.

To develop a fully decentralized method, we approach MARL in resource allocation as a distributed learning problem within a decentralized partially observable Markov decision process with individual rewards (Dec-POMDP-IR) framework. This framework accurately models the system dynamics in both conflict graphs and cellular networks. While a comprehensive theoretical study is beyond the scope of this work, we carefully refine the CTDE framework, adopting the multi-agent proximal policy optimization (MAPPO) algorithm with recurrent neural networks to propose two MARL-based solutions for the Dec-POMDP-IR problem. Our decentralized training and execution framework utilizes only local information during both training and execution phases, ensuring scalability. Extensive simulation results demonstrate the effectiveness and robustness of our proposed solutions across various network configurations.

One other key distinction of our work from previous studies~\cite{shi2011iteratively, shen2018fractional, huang2006distributed , kiani2007maximizing, zhao2021distributed, naderializadeh2014itlinq, sun2018learning,guo2020joint,nasir2019multi,tan2020deep,khan2020centralized,ge2023deep,chang2023federated} is the QoS metric.
Unlike prior works focusing on throughput maximization using sum-rate as the key performance metric, we prioritize average packet delay as QoS metric for two main reasons. First, wireless networks often operate under lighter traffic conditions than their maximum throughput capacity allows, making latency a more relevant measure of user experience. Second, high throughput does not necessarily eliminate significant packet delays, which can occur due to unbalanced scheduling that disproportionately favors certain links. Given this focus on delay, we treat varying queue length information as crucial and formulate a tractable delay minimization problem. We propose a traffic-driven MARL method for resource allocation, carefully designing the state, reward and transition based on queue information. Our work aims to learn flexible and adaptable policies that map dynamic traffic and CSI to a broad spectrum of actions. Unlike approaches that converge to static solutions~\cite{sun2018learning, zhao2021distributed, guo2020joint, nasir2019multi, tan2020deep, khan2020centralized, ge2023deep,chang2023federated}, our approach does not require learning entirely new policies when traffic conditions change. Instead, the neural network is designed to generalize across a range of traffic conditions, handling fluctuations in traffic loads and channel conditions without retraining.

This paper presents several key contributions:
\begin{itemize}
    \item We formulate traffic-driven resource allocation as a distributed learning problem within the Dec-POMDP-IR framework, incorporating partial observation, individual rewards, and local information sharing. We apply this formulation to conflict graphs and cellular networks, detailing the design of state spaces, action spaces, and reward functions.

    \item We adapt the conventional CTDE framework to decentralized training and execution, ensuring that both the training cost and neural network size remain constant for each agent, regardless of the system's scale. We implement recurrent neural networks and MAPPO in this process, presenting a detailed flow chart of the process and information exchange.

    \item We validate our solution's performance, scalability, and robustness through extensive simulations across various network configurations and traffic conditions. 

\end{itemize}

The paper is organized as follows: Section~\ref{sec:MARL framework and system model} formulates the learning problem and describes the conflict graph and cellular network systems. Section~\ref{sec: MARL solution} proposes two MARL-based solutions. Section~\ref{sec: simulation results} discusses the simulation setup and numerical results. Section~\ref{sec:Con} provides concluding remarks.

\section{MARL Framework and System Model}
\label{sec:MARL framework and system model}
\subsection{MARL framework}
\label{sec:MARL framework}
Before introducing the system model, we first establish the necessary reinforcement learning background. An {\em agent} is an entity that can process information from environment and make decisions to obtain desirable rewards. Consider multiple agents (indexed as $k \in \{1, 2, \ldots, K\}$) interacting with their environment over discrete time steps $t = 1, 2, \ldots, \mathcal{T}$, where $\mathcal{T}$ is the episode length.
At each time step $t$:
\begin{itemize}
    \item The environment is described by a global state $\mathbf{s}^{(t)}$, which contains the necessary variables that can accurately represent the dynamics of the environment. 
    \item Agent $k$ takes an action $a_k^{(t)}$ based on its {\em belief} of the environment, collectively forming a joint action $\mathbf{a}^{(t)} = \{a_1^{(t)}, \ldots, a_K^{(t)}\}$. The {\em belief} is derived on agent $k$'s local observation $O_k^{(t)}$ and historical observations $\tau_k^{(t)}$ from the environment. Notably, $O_k^{(t)}$ typically represents only a partial observation of the global state $\mathbf{s}^{(t)}$. In our setting, we introduce the concept of a neighborhood for each agent, comprising the agent itself and its neighboring agents. This setup allows agents to share their observations within their neighborhood, further enriching the {\em belief}.
    
    \item We assume a Markov transition model, the transition probability from the current global state $\mathbf{s}^{(t)}$ to the next global state $\mathbf{s}^{(t+1)}$ is solely determined by $\mathbf{s}^{(t)}$ and the current joint action $\mathbf{a}^{(t)}$, independent of the historical states and actions: 
    \begin{align} \label{eq: transition probability}
    p\left( \mathbf{s}^{(t+1)}|\mathbf{s}^{(t)},\mathbf{a}^{(t)} \right) .
    \end{align}
    \item At the end of each time step, agent $k$ receives an individual reward $R_k^{(t)}$. 
\end{itemize}

Building upon these elements and drawing inspiration from Dec-POMDP~\cite{oliehoek2016concise}, we model the multi-agent learning problem with individual reward as {\em Decentralized Partially Observable Markov Decision Process with Individual Rewards} (Dec-POMDP-IR). In this framework, $\mathcal{K}$ represents the set of agents. The state space, $\mathcal{S}$, encompasses all possible states, with $\mathbf{s}^{(t)} \in \mathcal{S}$. For each agent $k$, we denote the action space as $\mathcal{A}_k$, where $a_k^{(t)} \in \mathcal{A}_k$ represents the action taken by agent $k$ at time $t$. The joint action space is defined as $\mathcal{A} = \mathcal{A}_1 \times \dots \times \mathcal{A}_K$. The transition probability function $\mathcal{P} : \mathcal{S} \times \mathcal{A} \times \mathcal{S} \to [0,1]$ specifies the transition probability $p(\mathbf{s}^{(t+1)}|\mathbf{s}^{(t)},\mathbf{a}^{(t)})$ defined in \eqref{eq: transition probability}. We also define the observation space for agent $k$ as $\Omega_k$ and observation function as $\mathcal{O}$, 
which maps the state to the local observation $O_k \in \Omega_k$ for every $k\in\mathcal{K}$. 
The reward function is refined as $\mathcal{R} : \mathcal{S} \times \mathcal{A} \times \mathcal{S} \to \mathbb{R}^{K}$, indicating that each agent receives an individual reward instead of a shared common reward in each transition. With discount factor $\gamma$ balancing immediate and future rewards, our Dec-POMDP-IR can be represented by the tuple $\langle \mathcal{K}, \mathcal{S}, \{\mathcal{A}_i\}_{i \in \mathcal{K}}, \mathcal{P}, \mathcal{R}, \{\Omega_i\}_{i \in \mathcal{K}}, \mathcal{O}, \gamma \rangle$.

\begin{figure}
\centering
\includegraphics[width=.8\mywidth]{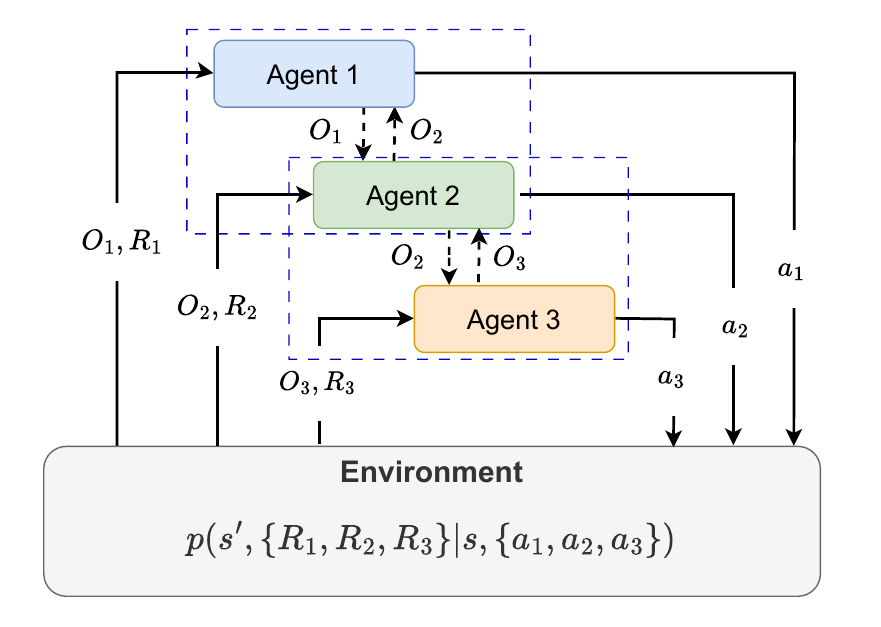}
\caption{Examples of Dec-POMDP-IR model with three agents.}
\label{fig: Dec-POMDP-IR}
\end{figure}

Fig.~\ref{fig: Dec-POMDP-IR} illustrates an example of agents-environment interaction in our framework. There are three agents, with Agents 1 and 2 forming one neighborhood, and Agents 2 and 3 forming another.
Agents receive local observations from the environment and communicate with their neighboring agent(s). Subsequently, agents make decisions based on these observations and their historical observations. The global state evolves based on joint actions and 
exogenous randomness, and the environment generates rewards for each agent for each state transition. 

The {\em policy} of agent $k$ is denoted as $\pi_k$, which represents a conditional probability distribution of actions based on the agent $k$'s belief. Agent $k$ samples its action $a_k$ from this distribution. The learning goal for agent $k$ is to find a good {\em policy} $\pi_k$ to maximize its own cumulative discounted reward: 
\begin{align} \label{eq: reward function}
    \mathbb{E}_{\pi} \left[\sum_{t=0}^{\infty} \gamma^{t}
    R_k^{(t)} \left(\mathbf{s}^{(t)}, \mathbf{a}^{(t)}, \mathbf{s}^{(t+1)}\right)\right].
\end{align}
where the expectation $\mathbb{E}_\pi$ assumes that the initial state is sampled from the 
initial state distribution, each agent follows its policy $\pi_k$ to select 
actions (i.e., $a_k^{(t)} \sim \pi_k(\cdot|{\text{ \em belief}}_k^{(t)})$), and that successor states are governed by the state 
transition probabilities (i.e., $\mathbf{s}^{(t+1)} \sim p(\cdot|\mathbf{s}^{(t)}, \mathbf{a}^{(t)})$). 
Notably, each agent's performance is influenced by both its own policy and those of other agents, emphasizing the importance of developing mutually beneficial policies for desirable rewards.

Our MARL problem formulation differs from other MARL problem formulations in resource allocation~\cite{nasir2019multi,tan2020deep,khan2020centralized,ge2023deep}. We do not define a local state space for each agent, and we do not assume {\em transition independence} across agents, as agents' actions (allocation decisions) significantly impact other agents' observations and
the evolution of the global state. The Markovian property in \eqref{eq: transition probability} holds for the global state and joint action but not for individual agents. Furthermore, we do not assume a common cooperative reward, as our goal is to develop a fully decentralized framework. 

In the remainder of this section, we present two concrete Dec-POMDP-IR models, for which we provide learning-based solutions in subsequent sections. The first model is simpler and the second model builds on the first one to describe a multi-cell wireless network with multiple frequency sub-bands.

\subsection{Conflict Graph}
\label{sec: conflict graph}
\subsubsection{System Model}

\begin{figure} 
\centering
\includegraphics[width=0.4\textwidth]{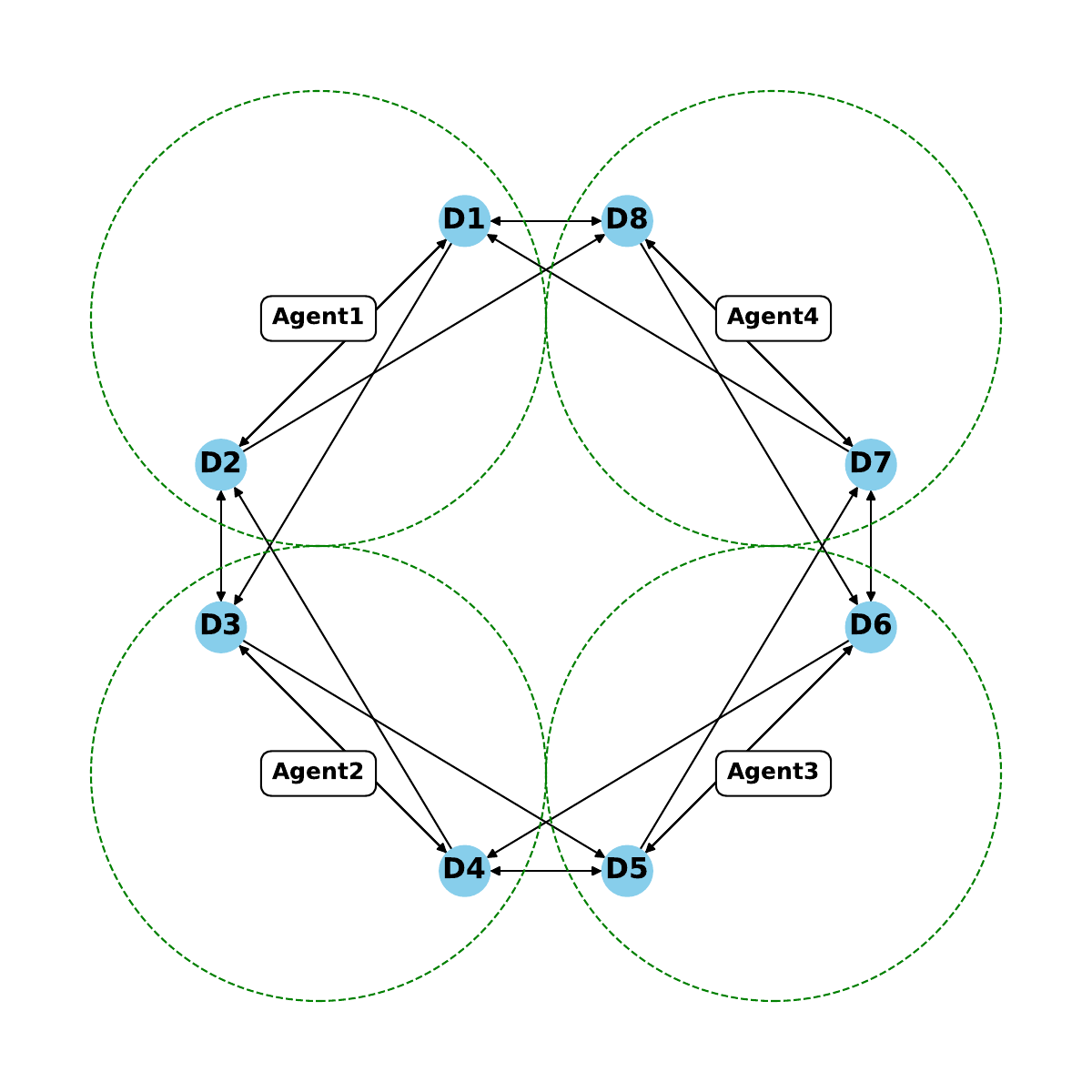}
\caption{A conflict graph of 4 agents in a symmetric deployment.}
\label{fig:8 link conflict graph}
\end{figure}

The challenge of scheduling in conflict graphs involves allocating resources (e.g., radio spectrum sub-bands, computing units) to conflicting tasks or events. Consider a directed conflict graph denoted as $\mathcal{G}=(\mathcal{I}, \mathcal{E})$, where each vertex in $\mathcal{I}$ represents a device, and an edge $(i,j)\in\mathcal{E}$ with $i,j\in\mathcal{I}$ and $i\ne j$ indicates device $i$ would cause conflict to device $j$ if they are scheduled simultaneously. 

Fig.~\ref{fig:8 link conflict graph} depicts a conflict graph where each agent serves two devices. Each device operates a first-in-first-out (FIFO) queue for assigned tasks. Time is slotted, and each device receives a random number of new tasks at the beginning of each time slot. For simplicity and without loss of generality, all tasks require identical resources to proceed and agents have unit capacity, meaning one task can be successfully processed during one time slot if the device is scheduled. Upon successful processing, the task departs from the queue.

The directional edges in Fig.~\ref{fig:8 link conflict graph} indicate conflicts between devices. For example, if device~1 is scheduled, it would potentially cause conflict with device~2, 3 and 8. We adopt the standard collision model, where a task is successfully processed if and only if no other conflicting devices are scheduled in the same time slot. If a conflict occurs, the task processing fails, and the task remains in the queue.

\subsubsection{Problem Formulation}
We now formulate the scheduling problem in conflict graph as a Dec-POMDP-IR model. In a $K$-agent $N$-device conflict graph, where $\mathcal{K}=\{1,2, \ldots, K\}$ and $\mathcal{N}=\{1,2, \ldots, N\}$ denote the set of agent indices and device indices, respectively. We define $b_n \in \mathcal{K}$ as the serving agent of device $n$. Consequently, $\mathcal{N}_{k}=\left\{n \in \mathcal{N} \mid b_{n}=k\right\}$ denotes all the devices served by agent $k$. In the example of Fig.~\ref{fig:8 link conflict graph}, $b_1=1$ and $\mathcal{N}_1=\{1,2\}$. 

At each time slot $t$, agent $k$ makes a scheduling decision $a_k^{(t)}$, which is selected from:
\begin{align} \label{eq: action space of conflict graph}
    \left\{0, 1, , \ldots, |\mathcal{N}_k|\right\} .
\end{align}
A decision of $0$ indicates that no device is scheduled by agent $k$ during time slot $t$, or alternatively, an agent may schedule one of its served devices.

To ensure that our design is fully distributed, selected devices are represented by their local indices under each agent's control. A local-to-global index mapping strategy is employed in the simulation to convert decisions from the local to global indices. We define a bijective function $f$ that maps local device and agent indices to global indices:
\begin{equation} \label{eq: local2global mapping}
f: \{(i,k) : 1 \leq i \leq |\mathcal{N}_k|, 1 \leq k \leq K\} \rightarrow \{1, \ldots, N\}
\end{equation}
where $(i,k)$ represents the $i$-th device controlled by agent $k$. The function $f$ maps a local index and cell index to its corresponding global index. For example, in Fig.~\ref{fig:8 link conflict graph}, $f(1,3)=5$ and $f(2,3)=6$.

Let $\mu_{n}^{(t)}$ and $m_{n}^{(t)}$ denote the scheduling decision and the number of successfully processed tasks of device $n$ in time slot $t$, respectively. The binary variable $\mu_{n}^{(t)} = 1$ indicates the device $n$ is scheduled in time slot $t$, which occurs when $f(a_{b_{n}}^{(t)}, b_{n}) = n$. Consequently, the number of successfully processed task, $m_{n}^{(t)}$, is determined as follows:
\begin{align}
m_{n}^{(t)} = \begin{cases}
1, & \text{if}~\mu_{n}^{(t)} = 1, \mu_{i}^{(t)} \neq 1~\text{for all}~ (i,n)\in\mathcal{E}\\
0, & \text{otherwise} .
\end{cases}
\end{align}
Specifically, $m_{n}^{(t)} = 1$ indicates that device $n$ is scheduled for conflict-free operation at time slot $t$. If there is a conflict or the device is not scheduled, then $m_{n}^{(t)} = 0$. 

\begin{figure*}
    \centering
    \includegraphics[width=.9\textwidth]{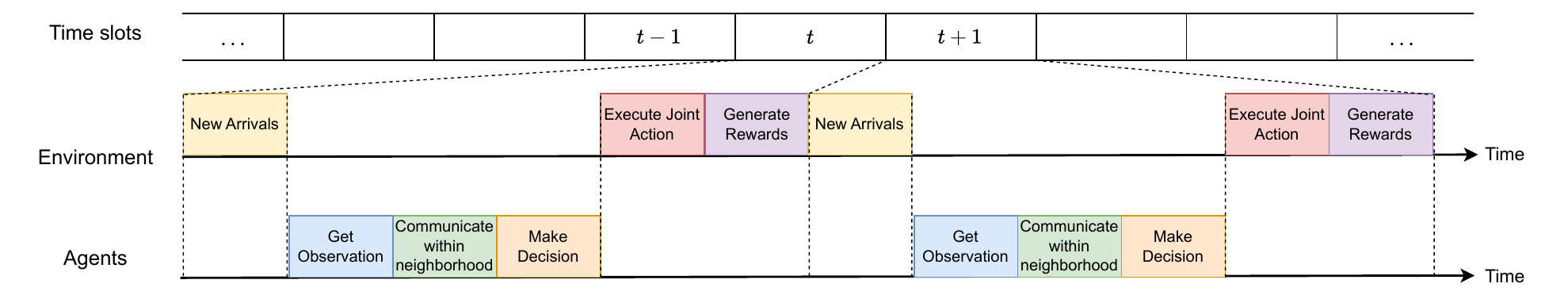}
    \caption{Illustration of the timing of interactions between agents and environments.}
    \label{fig: timeline}
\end{figure*}

To model the queueing dynamics for each device, let $Y_n^{(t)}$ denote the number of newly arrived tasks to device $n$ at the beginning of time slot $t$. The queue length of device $n$ at the end of slot $t$ can then be expressed as:
\begin{align} \label{eq: traffic dynamic conflict graph}
    q_{n}^{(t)}
    =
    \max \left(0, \, q_{n}^{(t-1)} + Y_{n}^{(t)} -  m_{n}^{(t)} \right).
\end{align}
We assume that $q_n^{(0)}=0$ since the queues start empty.

To better represent the system state, we introduce the queue length of device $n$ after receiving new packet arrivals as 
\begin{align}\label{eq: queue length entry conflict graph}
    \zeta_n^{(t)} = q_n^{(t-1)} + Y_n^{(t)}.
\end{align}
Given this, we can now define the global state $\mathbf{s}^{(t)} \in \mathcal{S}$ of the $K$-agent $N$-device conflict graph at each time slot $t$ as:
\begin{align}\label{eq: queue length entry} 
    \mathbf{s}^{(t)} = \left( \{q_{n}^{(t)}\}_{n=1}^N, \{\zeta_n^{(t)}\}_{n=1}^N \right)
\end{align}
This state representation encapsulates both the queue lengths after task processing and the updated queue lengths after new arrivals.

With the scheduling decision of the agent $k$, $a_k^{(t)}$, defined in \eqref{eq: action space of conflict graph}, we denote the joint action of all agents as $\mathbf{a}^{(t)} = \{a_1^{(t)}, \ldots, a_K^{(t)}\}$. Based on the traffic dynamic described in \eqref{eq: traffic dynamic conflict graph}, the transition from current global state $\mathbf{s}^{(t)}$ to next global state $\mathbf{s}^{(t+1)}$ is Markovian, allowing us to define Markov state transition model for $K$-agent $N$-device conflict graph:
\begin{align} 
p\left( \mathbf{s}^{(t+1)}|\mathbf{s}^{(t)},\mathbf{a}^{(t)} \right).
\end{align}

Next we discuss the accessible information for each agent, which forms the agent $k$'s {\em belief}. As mentioned in Section~\ref{sec:MARL framework}, each agent has a neighborhood, and we limit the information exchange within its neighborhood. Here we simply let agent $k$'s neighborhood be defined to include all agents whose devices conflict with those served by agent $k$.
For instance, in Fig.~\ref{fig:8 link conflict graph}, agent~1's neighborhood includes agents~2 and~4, while agent~2's neighborhood includes agents~1 and~3, and so on. Let $l_k$ denote the number of neighbors agent $k$ has, and let $\nu_{k,1},\dots,\nu_{k,l_k}$ denote their indexes. Let $\mathcal{C}(k)= \{k,\nu_{k,1},\dots,\nu_{k,l_k}\}$ denote agent $k$'s neighborhood, which always includes the agent itself. Agent $k$ utilizes information from $\mathcal{C}(k)$ to make scheduling decisions for all devices it served.

Using the local to global mapping $f$ defined in~\eqref{eq: local2global mapping},
$f(1,k),\dots,f(|N_k|,k)$ denote the global indexes of devices served by agent $k$. The local observation of agent $k$ at time slot $t$, denoted by $O_k^{(t)}$, is defined as:
\begin{align}\label{eq: state entry}
O_k^{(t)} = \left\{ \zeta_{f(1,k)}^{(t)}, \ldots, \zeta_{f(|\mathcal{N}_k|,k)}^{(t)} \right\},
\end{align}
which includes queue length information after new arrivals for all devices it serves. As information exchange within the neighborhood is beneficial for better agent inference, the local aggregate information of agent $k$ at time slot $t$, denoted by $X_k^{(t)}$, includes the local observations of agent $k$'s neighboring agents and itself.
\begin{align}\label{eq: aggregate information}
X_k^{(t)} = \left\{ O_k^{(t)}, O_{\nu_{k,1}}^{(t)}, \dots, O_{\nu_{k,l_k}}^{(t)} \right\},
\end{align}

Each agent maintains a local observation history $\tau_k$ for a time horizon $\Upsilon$. The observation history at time slot $t$ is defined as $\tau_k^{(t)} = \left(X_k^{(t-\Upsilon)}, \ldots, X_k^{(t - 1)} \right)$.

Since our goal is to minimize the delay, we define the learning objective using queue lengths as surrogates. Evidently, longer queue lengths lead to longer delays. Specifically, the direct contribution of agent $k$ to the queue length objective can be expressed as:
\begin{align}\label{eq: utility}
	u_{k}^{(t)} (\mathbf{s}^{(t)})= -\sum_{i \in \mathcal{N}_k} q_{i}^{(t)}.
\end{align}
To promote collaborative behavior and encourage joint decisions that lead to mutually beneficial outcomes, we also incorporate the utilities of agent $k$'s neighbors as indirect contributions. This approach discourages overly aggressive scheduling that might lead to frequent conflicts and performance degradation. Consequently, the individual reward function of agent $k$ is defined as:
\begin{align} \label{eq:reward function}
    R_k^{(t)}\left(\mathbf{s}^{(t)}\right)
    =\sum_{i \in \mathcal{C}(k)} u_i^{(t)}.
\end{align}
It is worth noting that we can also define the reward function $\mathcal{R}$ in a simpler version in this setting: $\mathcal{R}: \mathcal{S} \to \mathbb{R}^{K}$. Although actions are not explicitly defined in this simplified definition, we still need to take good actions that move the system to more favorable global states (i.e. shorter queue lengths).

A key feature of our design is that despite the reward $R_k$ generally depending on global states, it can be computed locally using only queue length information from agent $k$ and its neighbors. For example, the reward for agent $1$ in Fig.~\ref{fig:8 link conflict graph} in slot $t$ is equal to $-\big(q_1^{(t)}+q_2^{(t)}+q_3^{(t)}+q_4^{(t)}+q_7^{(t)}+q_8^{(t)}\big)$, which agent $1$ can compute using its own information and information from neighboring agents $2$ and $4$.

For clarity, we present the time flow of interactions between agents and environments in Fig.~\ref{fig: timeline}.

\subsection{Cellular network}
\subsubsection{system model}
\label{sec: SINR model}

\begin{figure}
\centering
\includegraphics[width=0.4\textwidth]{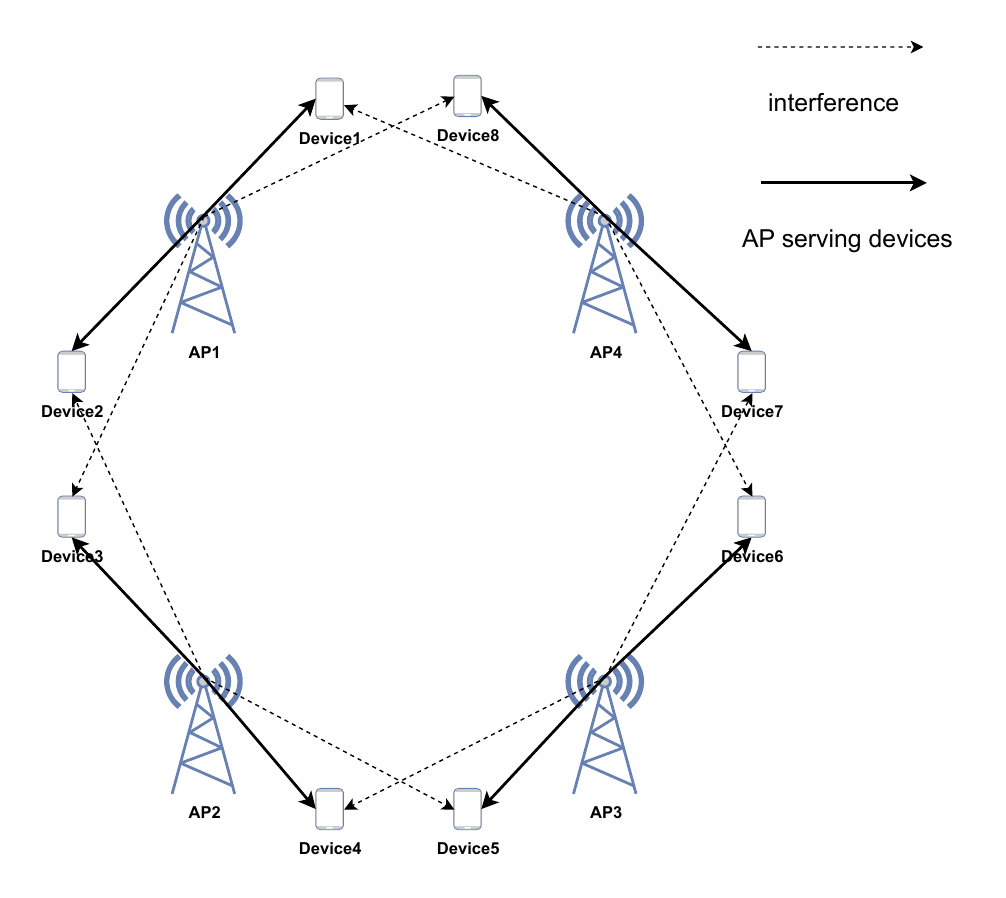}
\caption{A symmetric deployment with 4 APs and 8 devices.}
\label{fig:8 link deployment}
\end{figure}

Resource allocation in wireless communication network is a natural fit of Dec-POMDP-IR model. While the conflict graph provides an effective abstraction of wireless communication networks, 
cellular networks offer a more detailed and realistic model. In fact, the conflict graph discussed in Section \ref{sec: conflict graph} is a simplified representation of the cellular network deployment illustrated in Fig.~\ref{fig:8 link deployment}.

We consider downlink transmissions in a cellular network comprising $N$ (mobile) devices served by $K$ access points (AP), one AP per cell. All transmitters and receivers are equipped with a single antenna. As in the conflict graph model, $\mathcal{K}=\{1,2, \ldots, K\}$ and $\mathcal{N}=\{1,2, \ldots, N\}$ denote the set of cell indices and device indices, respectively. Each device $n \in \mathcal{N}$ is associated with its nearest AP, indexed as $b_n \in \mathcal{K}$. We refer to the downlink from $b_n$ to device $n$ as link $n$. The set of devices served by AP $k$ is denoted as $\mathcal{N}_{k}=\{n \in \mathcal{N} \mid b_{n}=k\}$.

Time is slotted with duration $T$, and the network utilizes $H$ orthogonal sub-bands. The downlink channel gain from transmitter $i$ to receiver $j$ in time slot $t$ on sub-band $h$ is expressed as:
\begin{align} \label{eq: downlink channel gain}
    g_{i\rightarrow j, h}^{(t)}=\alpha_{i\rightarrow j} \left|\beta_{i\rightarrow j, h}^{(t)}\right|^{2} , \quad t=1,2, \ldots
\end{align}
where $\alpha_{i\rightarrow j} \geq 0$ represents the large-scale path loss, which remains constant over many time slots. And $\beta_{i\rightarrow j, h}$ represents a small-scale Rayleigh fading component. In simulations, we use a first-order complex Gauss-Markov process to model small-scale fading:
\begin{align} \label{eq: small-scale fading}
    \beta_{i\rightarrow j, h}^{(t)}=\rho \beta_{i\rightarrow j, h}^{(t-1)}+\sqrt{1-\rho^{2}} e_{i\rightarrow j, h}^{(t)}
\end{align}
where $\big( \beta_{i\rightarrow j, h}^{(0)},
e_{i\rightarrow j, h}^{(1)}, e_{i\rightarrow j, h}^{(2)}, \ldots \big)$ are independent and identically distributed circularly symmetric complex Gaussian random variables with unit variance.

The power allocated to transmitter $n$ by its associated AP $b_n$ in time slot $t$ on sub-band $h$ is denoted as $p^{(t)}_{n, h}$. 
Assuming additive white Gaussian noise with power $\sigma^2$ for all receivers across all sub-bands, the downlink spectral efficiency of link $n$ in time slot $t$ on sub-band $h$ is:
\begin{align} \label{eq: spectral efficiency}
C_{n,h}^{(t)}= \log\left(1 + \frac{g_{n\rightarrow n, h}^{(t)}~p^{(t)}_{n, h}}{\sum_{j \in \mathcal{N}, j \neq n} g_{j\rightarrow n,h}^{(t)}~p^{(t)}_{j , h}+\sigma^{2}}\right).
\end{align}

Each AP acts as an agent, scheduling transmissions and allocating power for all devices within its cell. The neighborhood concept applies here as well, with agent $k$'s neighborhood including all agents whose devices may cause sufficiently high interference to the devices in $\mathcal{N}_k$. Specifically, if the pathloss component $\alpha_{b_n \rightarrow n} - \alpha_{k \rightarrow n}$ falls below a certain threshold, the device $n$ is considered to be potentially highly interfered by devices in $\mathcal{N}_k$. Consequently, the neighborhood of agent $k$ would include agent $b_n$. 

For practical reasons, an AP cannot serve multiple links on the same sub-band simultaneously. In each time slot, the agent $k$ needs to make scheduling decision and decide the transmission power $p_{k,h}^{(t)}$ for on each sub-band $h$:
\begin{align} \label{eq: action space}
z_{k,h}^{(t)} &\in \left\{0, 1, , \ldots, |\mathcal{N}_k|\right\} , \\
p_{k,h}^{(t)} &\in \left\{P_{\min }, P_{\min }\left(\frac{P_{\max }}{P_{\min }}\right)^{\frac{1}{\left|\mathcal{P}\right|-1}}, \ldots, P_{\max }\right\} .
\end{align}
A decision of $z_{k,h}^{(t)}=0$ indicates that no links in cell $k$ are activated during time slot $t$ on sub-band $h$. Alternatively, an agent may select one of the links within its cell for transmission using transmission power from a quantized log-step power options ranging from $P_{\min}$ to power constraint $P_{\max}$. Links that are not selected remain silent in time slot $t$ on sub-band $h$ (i.e., power set to 0). 

Without loss of generality, we assume identical packet size. Let $L$ denotes the packet size in bits, and $W_h$ denotes the bandwidth of sub-band $h$. The queueing dynamics for each link with the queue length (in bits) of link $n$ at the end of slot $t$ expressed as:
\begin{align} \label{eq: traffic dynamic}
    q_{n}^{(t)}
    =\max \left(0, \, q_{n}^{(t-1)} + Y_{n}^{(t)} L - \sum_{h=1}^H C_{n,h}^{(t)} W_h T \right)
\end{align}
where $Y_n^{(t)}$ denotes the number of newly arrived packets to device $n$ at the beginning of time slot $t$. The spectral efficiency is a function of decision variable $\left( z_{b_n,h}^{(t)}, p_{b_n,h}^{(t)}\right)$. 

\subsubsection{problem formulation}
The cellular network system described can be formulated as Dec-POMDP-IR as well. Define the cellular network CSI at time slot $t$ as a $N \times N \times H$ tensor as:
\begin{align} \label{eq: channgel gain tensor}
\mathbf{G}^{(t)} = \left\{ g_{i\rightarrow j, h}^{(t)} \mid i,j \in \{1, \ldots, N\}, h \in \{1, \ldots, H\} \right\}
\end{align}
which represents the channel gain between all transmitters and receivers across all sub-bands. 
At each time slot $t$, the global state $\mathbf{s}^{(t)} \in \mathcal{S}$ of the $K$-agent $N$-device cellular network is given by:
\begin{align}\label{eq: queue length entry} 
    \mathbf{s}^{(t)} = \left( \mathbf{G}^{(t)}, \{q_{n}^{(t)}\}_{n=1}^N, \{\zeta_n^{(t)}\}_{n=1}^N \right)
\end{align}
where $\zeta_n^{(t)} = q_n^{(t-1)} + Y_n^{(t)} L$ represents the queue length of device $n$ after receiving new packet arrivals (where agents measure/get observation from environments as shown in Fig.~\ref{fig: timeline}.

The action of agent $k$ at time slot $t$ is defined as $a_k^{(t)} = \left( z_{k,h}^{(t)}, p_{k,h}^{(t)}\right)_{h=1}^H \in \mathcal{A}_k$, indicating the device selection and corresponding power level across all sub-bands.
The Markov state transition model for the cellular network system follows that of the conflict graph, with the Markovian transition probability from the current global state $\mathbf{s}^{(t)}$ to the next global state $\mathbf{s}^{(t+1)}$ defined as $p\left( \mathbf{s}^{(t+1)}|\mathbf{s}^{(t)},\mathbf{a}^{(t)} \right)$.

The accessible information for each agent in the cellular network model is more detailed than in the conflict graph model. We assume that for each device $n$, the transmitter learns the direct channel gain $g_{n, h}$ on sub-band $h$ via receiver feedback, while the receiver measures the total interference-plus-noise power and its spectral efficiency. Both transmitters and receivers report the CSI information to the corresponding agent (cell) $b_n$, but the CSI information is delayed by one time slot. The transmitter also records the transmission power from previous time slot. Additionally, we assume that each agent has timely queue length information for all links within its cell. Therefore, the accessible information $o_n^{(t)}$ for device $n$ at time slot $t$ includes:
\begin{itemize}
\item $\zeta_n^{(t)}$: the queue length of device $n$
\item $\left\{g_{n\rightarrow n, h}^{(t-1)}\right\}_{h = 1}^{H}$: the direct gain on each sub-band
\item $\left\{p^{(t-1)}_{n, h}\right\}_{h = 1}^{H}$: device $n$'s action decision on each sub-band
\item $\left\{\sum_{j \in \mathcal{N}, j \neq n}g_{j\rightarrow n,h}^{(t-1)}~p^{(t-1)}_{j, h}+\sigma^{2}\right\}_{h = 1}^{H}$: the interference-plus-noise power at receiver $n$ on each sub-band
\item $\left\{C_{n,h}^{(t-1)}\right\}_{h = 1}^{H}$: spectral efficiency of link $n$ computed from~\eqref{eq: spectral efficiency} on each sub-band
\end{itemize}

As in the conflict graph, we employ a local-to-global index mapping strategy to ensure a fully distributed design. The local observation of agent $k$ at time slot $t$, denoted by $O_k^{(t)}$, is defined as:
\begin{align}\label{eq: state entry}
O_k^{(t)} = \left\{ o_{f(1,k)}^{(t)}, \ldots, o_{f(|\mathcal{N}_k|,k)}^{(t)} \right\},
\end{align}
which includes delayed CSI information, delayed action and timely queue length information of all links within its cell. 
Then the local aggregate information of agent $k$ at time slot $t$ is $X_k^{(t)} = \left\{ O_k^{(t)}, O_{\nu_{k,1}}^{(t)}, \dots, O_{\nu_{k,l_k}}^{(t)} \right\}$ and the observation history at time slot $t$ is$\tau_k^{(t)} = \left(X_k^{(t-\Upsilon)}, \ldots, X_k^{(t - 1)} \right)$.

The reward function for the cellular network model is defined analogously to that of the conflict graph model, aiming to minimize packet delay using queue lengths as surrogates. The direct and indirect contributions to the queue length objective, as well as the individual reward function for each agent, are calculated using the same formulations presented in \eqref{eq: utility} and \eqref{eq:reward function} of the conflict graph model.

This formulation of our cellular network system and conflict graph as a Dec-POMDP-IR captures the essence of decentralized decision-making under partial observability and constrained information sharing, with individual rewards for each agent. The primary objective for each agent $k$ is to devise an optimal policy $\pi_k$ that effectively maps the local aggregate information $X_k$ in each time slot to a strategic sequence of actions $a_k$. This policy serves a dual purpose: 1) to maximize the agent's own reward function, as defined in equation \eqref{eq:reward function}, and 2) as a consequence, to minimize overall packet delay within the network.

\section{MARL-based solution}
\label{sec: MARL solution}
This section presents our MARL-based solution to the Dec-POMDP-IR problem formulated earlier. We aim to find good control policies that yield desirable rewards within the constraints of decentralized decision-making and partial observability. The policy $\pi_{k}$ of agent $k$ is determined by a policy network parameterized by $\theta_k$ (denoted as $\pi_{\theta_k}$ in this section), which maps the local aggregate information $X_k^{(t)}$ and its history $\tau_k^{(t)}$ to a categorical distribution over discrete actions. The value network, parameterized by $\phi_k$, estimates the expected return from a given state based on local aggregate information $X_k^{(t)}$ and its history $\tau_k^{(t)}$. 

The policies are trained in parallel using trajectories of states, actions, and rewards. Our approach refines the popular on-policy training algorithm MAPPO, which has demonstrated success in various cooperative multi-agent tasks \cite{yu2022surprising}. We adapt this algorithm to our specific setting and incorporate recurrent neural networks to process historical information effectively. Based on this framework, we propose two distinct training methods, each with its own strengths and trade-offs. First we describe the important recurrent neural network structures in the network.

\subsection{Recurrent Neural Network}

We incorporate recurrent neural network structures, specifically long short-term memory (LSTM) units, into both the policy and value networks for the following reasons: 1) The state transition for each individual agent is non-Markovian, and making decisions based on information from a single time step is insufficient due to partial observability. 2) To make decisions based on historical information, directly inputting all historical data $\tau_k^{(t)}$ can be redundant and increase network size. LSTM layers can carry important information through the cell state and discard redundant information.

Both the policy network and value network contain two parts: an LSTM layer for history embedding and a multi-layer perceptron (MLP) 
for decision making/value estimation. We define the recurrent state of the policy network for agent $k$ at time slot $t$ to be $\hat{X}_k^{(t)}$, which serves as a compact representation of the history $\tau_k^{(t)}$. Similarly, we define the recurrent state of the value network for agent $k$ at time slot $t$ as $\tilde{X}_k^{(t)}$.

By utilizing a recurrent architecture, agents can capture temporal dependencies and adapt to environment dynamics beyond a single observation, which allows agents to better infer neighboring agents' behaviors and impacts, making informed decisions based on augmented context. This sequential memory approach also encourages consideration of both immediate and long-term effects in the decision-making process.

For the policy network of agent $k$ at time slot $t$, the input includes local aggregate information $X_k^{(t)}$ and the previous time slot's recurrent state $\hat{X}_k^{(t-1)}$. The output includes the recurrent state of this time slot $\hat{X}_k^{(t)}$ (generated by the LSTM layer) and the action decision (generated by the MLP). The recurrent state is updated iteratively across time steps and carries important information as a historical embedding. The agent makes decisions based on the current time slot's local aggregate information and this embedding. Similarly, the value network takes $X_k^{(t)}$ and $\tilde{X}_k^{(t-1)}$ as input and outputs a value estimation and $\tilde{X}_k^{(t)}$.

To ease implementation, we introduce dummy links to maintain identical state and action space dimensions for all agents, regardless of the number of devices they serve, under the practical assumption this number is capped by a constant. Next we discuss specifics of two MARL-based solutions.

\subsection{Individual Policies}
\label{sec: individual policies}
Our first method implements a fully distributed approach for both training and execution. Inspired by the scalable framework in \cite{qu2020scalable}, we modify the typical CTDE process. Each agent $k$ maintains its own policy and value networks and both input only local information $X_k^{(t)}$ and recurrent state. Specifically, the input to policy network is defined as $\hat{\mathcal{X}}_k^{(t)} = \{X_k^{(t)}, \hat{X}_k^{(t-1)}\}$, the input to policy network is defined as $\tilde{\mathcal{X}}_k^{(t)} = \{X_k^{(t)}, \tilde{X}_k^{(t-1)}\}$. By limiting the neighborhood size, we ensure that network input dimensions remain constant regardless of the total number of agents in the system. This design enables truly decentralized operations, as each agent operate independently, managing its own trajectory, sampling from it, and training its networks to maximize its individual reward $R_k$. The resulting method is highly scalable and practical for large-scale networks.

For simplicity and formula reusability, we describe the decentralized training and execution process for agent $k$ without carrying the sub-index $k$ in the following formulas. Throughout this subsection, the reward $R$, policy network $\theta$, value network $\phi$, policy network input $\hat{\mathcal{X}}$, value network input $\tilde{\mathcal{X}}$, action $a$ and sample batch $B$ refer to the corresponding variables of agent $k$.

The training process is iterative, with both the policy and value networks being updated for a fixed number of steps after each episode. The networks from the previous training step are denoted as $\theta_{\text{old}}$ and $\phi_{\text{old}}$. We first estimate the advantage function by the truncated version of generalized advantage estimation (GAE) in~\cite[Eq.~16]{schulman2015high} based on episode trajectory, for each time slot $t$:
\begin{align} \label{eq: advantage function}
A^{(t)} = \sum_{l=0}^{\mathcal{T}-t-1} (\gamma \lambda)^l \left(R^{(t+l)} + \gamma V_{\phi}(\tilde{\mathcal{X}}^{(t+l+1)}) - V_{\phi}(\tilde{\mathcal{X}}^{(t+l)})\right)
\end{align}
where $\lambda$ is the exponentially-weighted hyper-parameter, $\mathcal{T}$ is the horizon of one episode and $A^{(\mathcal{T})} = V_{\phi}(\tilde{\mathcal{X}}^{(\mathcal{T})})$ as special case.

After computing the advantage function for all time slots in the trajectory, we sample a batch of transitions from the trajectory with size $|B|$, where $B$ stands for the sample of time indexes. The sampled policy network input $\hat{\mathcal{X}}$, value network input $\tilde{\mathcal{X}}$, actions $a$ and corresponding advantages $A$ are used to update the networks. The value network parameters are updated to fit the estimated advantage values by minimizing the following loss function:
\begin{align} \label{eq: critic loss function}
\mathcal{L}(\phi, \tilde{\mathcal{X}}, A)
&= \frac{1}{|B|}  \sum_{t \in B}
\max \Bigg[\left(V_\phi\left(\tilde{\mathcal{X}}^{(t)}\right)-A^{(t)} \right)^2, \nonumber \\
& \quad \left(c_\epsilon\left(V_\phi\left(\tilde{\mathcal{X}}^{(t)}\right) , V_{\phi_{\text{old}}}\left(\tilde{\mathcal{X}}^{(t)}\right)\right)
-A^{(t)}\right)^2\Bigg]
\end{align}
where 
\begin{align}
  c_\epsilon(x,y) = \min( \max(x,y-\epsilon), y+\epsilon)
\end{align}
is a clipping function. 

We define the probability ratio: 
\begin{align}
    r_\theta(\hat{\mathcal{X}}, a)
    ={\pi_\theta\left(a \mid \hat{\mathcal{X}}\right)} \bigg/ {\pi_{\theta_{\text{old}}}\left(a \mid \hat{\mathcal{X}}\right)}. 
\end{align}
Let $\mathcal{H}(\cdot)$ denote the Shannon entropy of a probability mass function and $\delta$ to be the entropy coefficient hyper-parameter. We update the policy network of agent $k$ to maximize the objective function:
\begin{align} \label{eq: policy network objective function}
J(\theta, \hat{\mathcal{X}}, A) =
&\frac{1}{|B|} \sum_{t \in B} \min \left(r_{\theta}^{(t)} A^{(t)},
c_\epsilon\ \left(r_{\theta}^{(t)}, 1 \right)
A^{(t)}\right)  \nonumber \\
&+ \delta \frac{1}{B}\sum_{t=1}^B \mathcal{H}\left(\pi_\theta\left(\,\cdot\,|\hat{\mathcal{X}}^{(t)}\right)\right) .
\end{align}

The combined policy and value networks constitute an actor-critic architecture, which generally enhances sample efficiency and accelerates convergence compared to actor-only (e.g., policy gradient) or critic-only (e.g., Q-learning) methods. In stochastic environments, trajectories can yield varying returns (defined as the discounted sum of future rewards from a given state), resulting in high variance when using returns directly as an objective function for policy network. While increasing batch size can mitigate this variance, it compromises sample efficiency. The value network, employing temporal difference learning for bootstrapping as evidenced in~\eqref{eq: advantage function}, provides more accurate return estimates. This approach reduces variance in the advantage function $A^{(t)}$, which is central to both the value network loss function in~\eqref{eq: critic loss function} and the policy network objective function in~\eqref{eq: policy network objective function}. Consequently, this formulation facilitates faster convergence and improved stability in the learning process.

\subsection{Decentralized Training and Execution}
\begin{figure*}
    \centering
    \includegraphics[width=.8\textwidth]{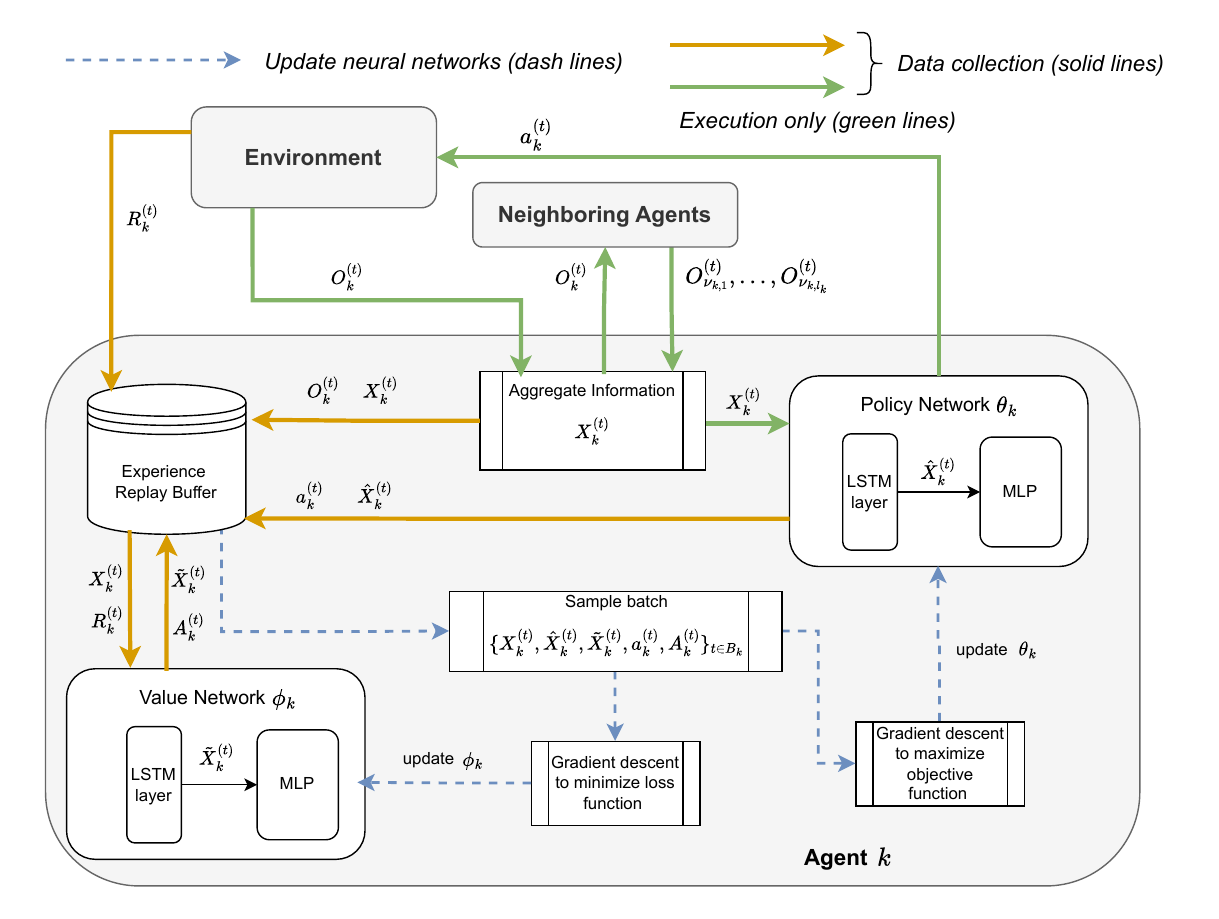}
    \caption{Diagram of the training and execution workflow.}
    \label{fig:training workflow}
\end{figure*}

Details of training process for agents with individual policies is summarized in Algorithm~\ref{alg:separate policy} and illustrated in Fig.~\ref{fig:training workflow}. The training procedure comprises two phases: data collection (indicated by solid lines in Fig.~\ref{fig:training workflow}) and networks update (indicated by dash lines in Fig.~\ref{fig:training workflow}). During the data collection phase, each agent $k$ operates under its current policy for an episode. Throughout this episode, the agent $k$ communicate local observation $O_k^{(t)}$ with neighboring agents and aggregates local information $X_k^{(t)}$. Provided with recurrent state from previous time slot, agent executes actions based on its policy $\pi_{\theta_k}\left(a_k^{(t)}|\hat{\mathcal{X}}_k^{(t)}\right)$. The agent records transitions, including local aggregate information $X_k^{(t)}$, recurrent state $\hat{X}_k^{(t)}$ (together forming input to policy network $\hat{\mathcal{X}}_k^{(t)}$), actions $a_k^{(t)}$, and rewards $R_k^{(t)}$. Subsequently, it computes advantage values $A_k^{(t)}$ retrospectively for each time step using Equation~\eqref{eq: advantage function} and records value network recurrent state $\tilde{X}_k^{(t)}$ (which, with local aggregate information, forms input to value network $\tilde{\mathcal{X}}_k^{(t)}$). All this information is then stored in the agent's experience replay buffer. 

The network update phase involves iterative refinement of the agents' policy and value networks. Each agent samples batch data $\big\{
{X}_k^{(t)}, \hat{\mathcal{X}}_k^{(t)}, \tilde{\mathcal{X}}_k^{(t)}, a_k^{(t)}, A_k^{(t)}\big\}_
{t \in B_k}$ 
from its replay buffer. The value network parameters $\phi_k$ are updated to minimize the critic loss function defined in Equation~\eqref{eq: critic loss function}, while the policy network parameters $\theta_k$ are adjusted to maximize the objective function given in Equation~\eqref{eq: policy network objective function}. This process is iterated for a predetermined number of episodes and iterations, facilitating policy improvement based on accumulated experience. Once the training is finished, only the policy network is employed during execution (indicated by green lines in Fig.~\ref{fig:training workflow}). This design ensures decentralized training and execution.

\begin{algorithm}
\caption{Decentralized training for agent $k$.}
\label{alg:separate policy}
\begin{algorithmic}[1]
\State Initiate policy network $\theta_k$ and value network $\phi_k$, initialize recurrent state $\hat{X}_k^{(0)}, \tilde{X}_k^{(0)}$
\For{each episode $e = 1,2,\ldots,E$}
    \State {/* Interact with environment and collect data */} 
    \For{time slot $t = 1,2,\ldots,\mathcal{T}$} 
        \State Communicate local information $O_k^{(t)}$ with neighboring agents.
        \State Take action based on $\pi_{\theta_k}\left(a_k^{(t)}|X_k^{(t)}, \hat{X}_k^{(t-1)}\right)$
        \State Record $\left(X_k^{(t)}, \hat{X}_k^{(t)}, a_k^{(t)}, R_k^{(t)}\right)$ to experience replay buffer.
    \EndFor
    \For{time slot $t = 1,2,\ldots,\mathcal{T}$} 
        \State Compute advantages $A_k^{(t)}$ using \eqref{eq: advantage function}.
        \State Record $\left(A_k^{(t)}, \tilde{X}_k^{(t)}\right)$ in experience replay buffer.
    \EndFor
    \State {/* Update policy and value networks */}
    \For{iteration $n = 1,2,\ldots,N_{iteration}$}
        \State $\phi_{k, old} \leftarrow \phi_{k}, \theta_{k, old} \leftarrow \theta_{k}$
        \State 
        Take $\{X_k^{(i)}, \hat{X}_k^{(i)}, \tilde{X}_k^{(i)}, a_k^{(i)}, A_k^{(i)}\}_{i \in B_k}$ as a sample batch from the experience replay buffer.
        \State Update $\phi_k$ to minimize \eqref{eq: critic loss function}.
        \State Update $\theta_k$ to maximize \eqref{eq: policy network objective function}.
    \EndFor
\EndFor
\end{algorithmic}
\end{algorithm}
\subsection{Shared Policy}
\label{sec: shared policies}
The second method employs a partially decentralized framework. While both policy and value networks still use only local information $\hat{X}_k$, all agents share a common policy and value networks, and optimize a common collective reward using shared trajectories. Compared with first method, this approach allows the shared policy to benefit from the experiences of all agents during training, and it is more effective when computation resource is limited.

The training process is similar to that of the individual policies method, but with the loss function for shared critic function be 
\begin{align} \label{eq: critic loss function centralized}
    \mathcal{L}'(\phi, \hat{X}, A) &= \frac1K \sum^K_{k=1} L(\phi, \hat{X}_k, A_k) 
\end{align}
and the objective function for shared policy network be:
\begin{align} \label{eq: policy objective function centralized}
    J'(\theta, \hat{X}, A) &= \frac1K \sum^K_{k=1} J(\theta, \hat{X}_k, A_k) .
\end{align}


\section{Simulation Results and Analysis}
\label{sec: simulation results}
\subsection{Simulation Setup}
\begin{figure}
\centering
\begin{subfigure}{0.9\columnwidth}
\centering
\includegraphics[width=\textwidth]{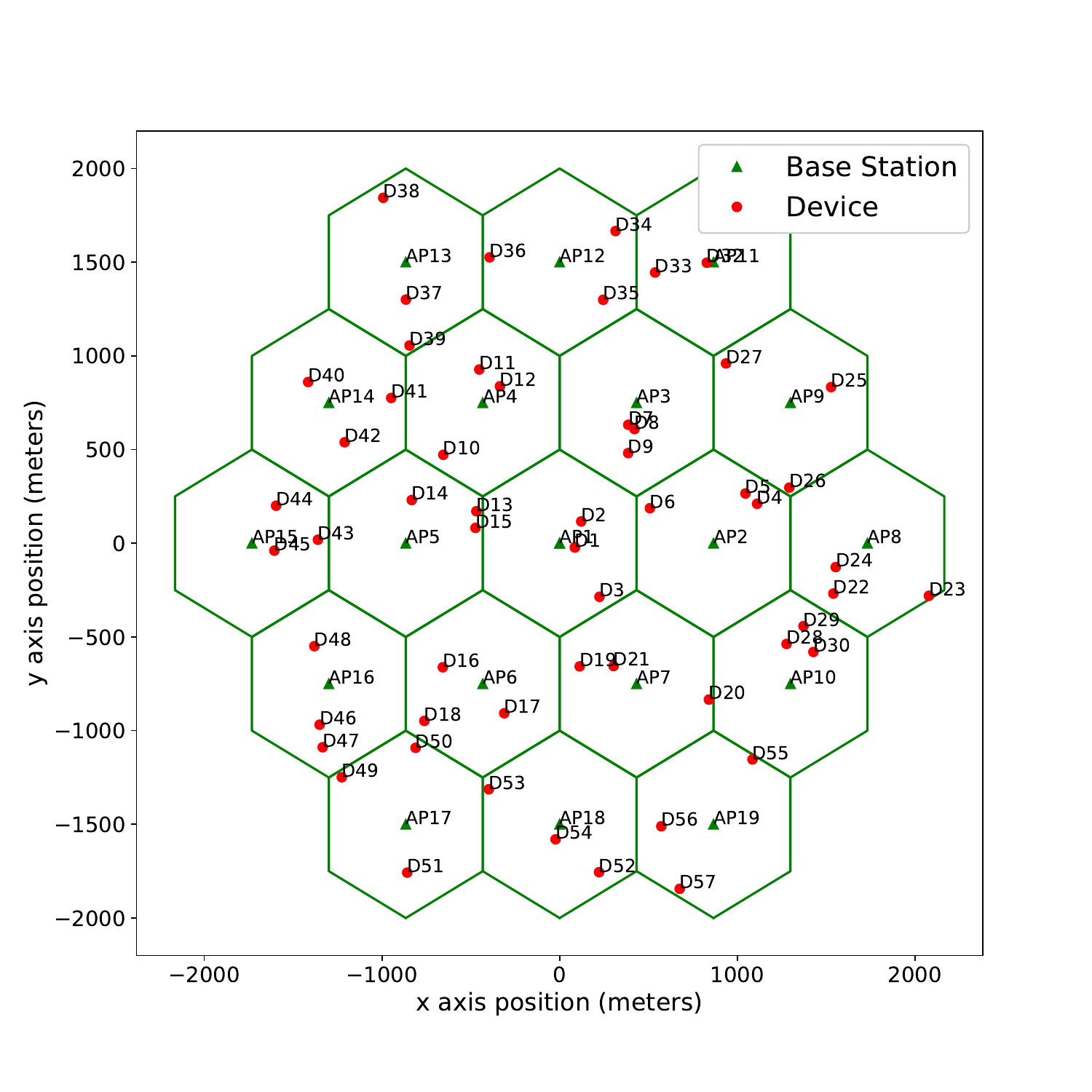}
\caption{}
\label{fig:57link hexa deployment}
\end{subfigure}%
\vspace{-0.4ex}
\
\begin{subfigure}{0.9\columnwidth}
\centering
\includegraphics[width=\textwidth]{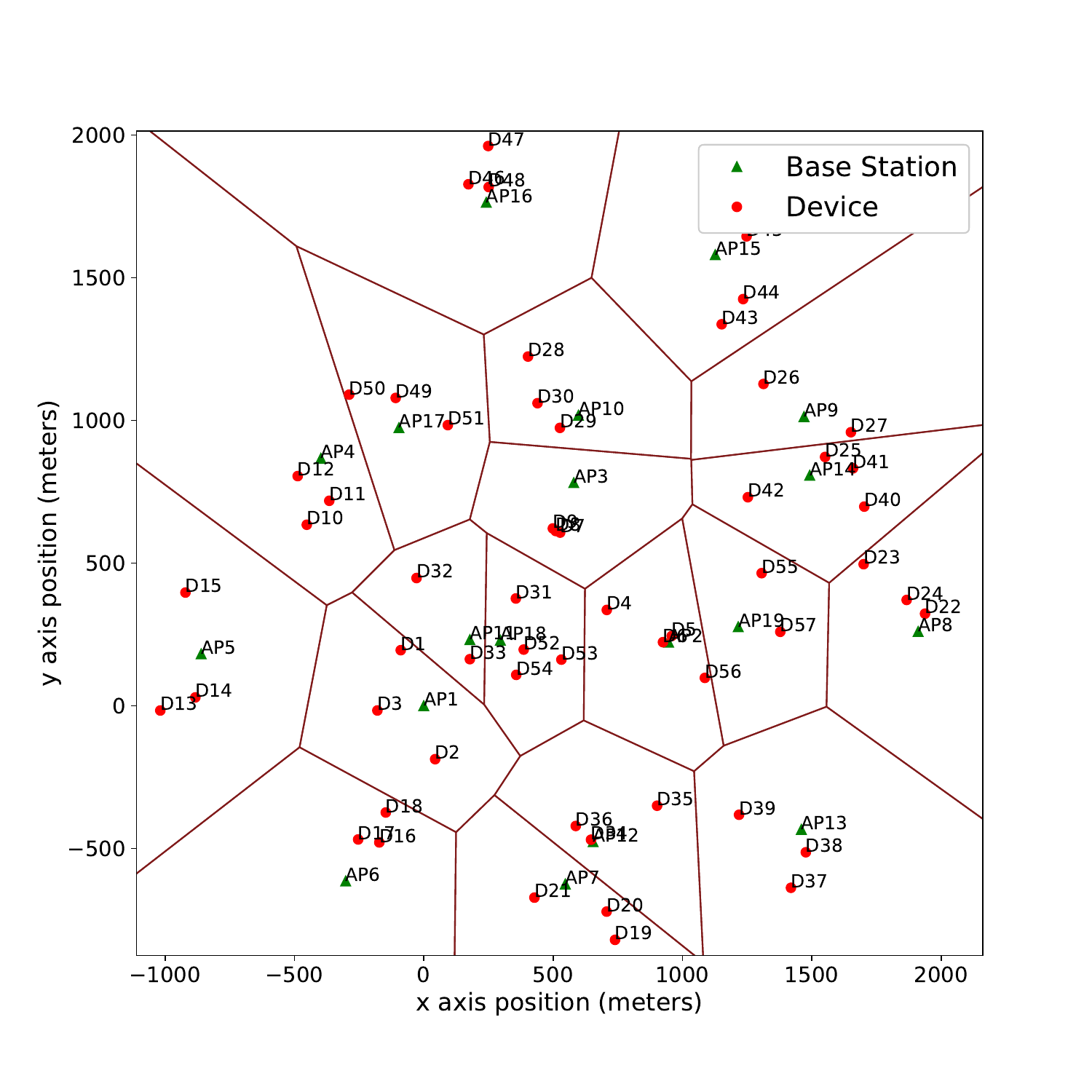}
\caption{}
\label{fig:57link random deployment}
\end{subfigure}
\caption{Two networks with 19 cells serving 57 devices. (a)~A regular 
deployment; 
(b) a random deployment.}
\label{fig:57link-deployment-combined}
\end{figure}

To evaluate the performance of proposed methods, we conducted simulations on both conflict graph and cellular network models under varying traffic intensities. In all scenarios, we let the number of packet arrivals to agent $n$ in time slot $t$, denoted by $Y_n^{(t)}$, be an independent Poisson random variable with mean $\lambda_n$. Throughout this section, we assume the spectrum is divided into $H=3$ sub-bands. Three distinct network configurations are considered:
\begin{enumerate}
    \item The conflict graph depicted by Fig.~\ref{fig:8 link conflict graph}, which is an abstraction of the downlink of the 8-device, 4-AP symmetrical deployment in Fig.~\ref{fig:8 link deployment}.
    \item A cellular network deployed as compact hexagons, consisting of 19 APs, with each AP serving 3 devices, as depicted in Fig.~\ref{fig:57link hexa deployment}.
    \item A randomly deployed cellular network with 19 APs and 57 devices, as illustrated in Fig.~\ref{fig:57link random deployment}. Each device is associated to its nearest AP. An AP servers between 2 and 5 devices.
\end{enumerate}
The cellular network simulations were conducted using the parameters listed in Table~\ref{tab:system_parameter}.

\begin{table} 
\centering
\begin{tabular}{|c|c|}
\hline
Cell radius: & 500m\\ \hline
Path loss (LTE standard): & $128.1 + 37.6 \log_{10}(\text{distance})$ (dB)\\  \hline
AWGN power: & $\sigma^2 = -114$ dBm\\ \hline
Max transmitter power: & $P_{max} = 23$ dBm\\ \hline
Discretized power levels: & $\left|\mathcal{P}\right| = 6$\\ \hline
Time slot duration: & $T = 20$ ms\\ \hline
Bandwidth for each sub-band: & $W_h = 20$ MHz\\ \hline
packet length& $L=0.5$ Mbits \\ \hline
\end{tabular}
\caption{Cellular network parameters}
\label{tab:system_parameter}
\end{table}

To comprehensively evaluate our proposed MARL-based scheduler, we compared its QoS performance against several benchmark schemes across both the conflict graph and cellular network scenarios. For the conflict graph setting, we employed three benchmark schemes:
\begin{enumerate}
    \item \textbf{GMS:} A centralized method that starts with an empty schedule, iteratively selects device with the longest queue in the network, adding it to the schedule and disabling those conflicting devices. This process repeats among the remaining devices until all devices are either scheduled or disabled.
     
    
    \item \textbf{LLQ:} A distributed greedy method in which each AP schedules a device for transmission if it has a longer queue than all devices it has a conflict with; in case of a tie between $j$ devices, each of those devices is scheduled independently with probability $1/j$. 
    
    \item \textbf{Q-CSMA~\cite{ni2011q}:} A method where devices perform carrier sensing prior to transmission, ensuring all scheduled devices form an independent set, and then each enabled device transmits with a certain probability based on its queue length.
 \end{enumerate}
 
For the cellular network scenarios, we utilized four benchmark schemes:
\begin{enumerate}
    \item \textbf{LLQ:} A greedy method where APs use all sub-bands at full power to serve the device with the longest queue in their neighborhood. In case of a tie between multiple devices, one device is selected uniformly at random.
    
    \item \textbf{ITLinQ~\cite{naderializadeh2014itlinq}:} The APs use full power and all sub-bands to serve subsets of devices with ``sufficiently'' low interference between them based on the CSI. 
    We actually simulate a slightly more complex version called Fair-ITLinQ~\cite{naderializadeh2014itlinq}, as the original ITLinQ exhibits poor performance in the 57-device scenario.

    \item \textbf{FP~\cite{shen2018fractional}:} An centralized iterative method based on minorization-maximization, assuming real-time global CSI is available. Device weights are proportional to queue lengths, and the sub-bands are allocated independently based on their respective CSI. To the best of our knowledge, the genie-aided FP method is essentially the best-performing resource allocation scheme, which performs similarly or outperform competitive techniques reported in~\cite{sun2018learning, nasir2019multi, guo2020joint, khan2020centralized, nasir2021deep, ge2023deep}.

    \item \textbf{WMMSE~\cite{shi2011iteratively}:} A centralized iterative optimization algorithm, also assuming real-time global CSI availability. Device weights are proportional to queue lengths, and the sub-bands are allocated independently based on CSI. Like FP, WMMSE is guaranteed to converge to a local optimum of the problem and offers comparable performance.
 \end{enumerate}

\subsection{Training and Testing}

We implemented both centrally trained shared policy and individually trained separate policies in Sec.~\ref{sec: individual policies} and \ref{sec: shared policies}. Each training episode lasted 2,000 time slots. 
To prevent the scheduler from being trapped in adverse queueing conditions before adequate training, episodes were terminated and restarted if any device's queue length exceeded a predefined threshold. For testing or deployment, episodes spanned 5,000 time slots, with th average packet delay serving as the performance metric when the queue is considered stable. Other MARL learning parameters are summarized in Table~\ref{tab:RL_parameter}.

\begin{table} 
\centering
\begin{tabular}{|c|c|}
\hline
Network optimizer & RMSprop for all neural networks\\  \hline
Learning rates & 0.0001 for all neural networks\\  \hline
Number of recurrent layers & 1 for all neural networks \\ \hline
Number of hidden layers & 2 for all neural network\\ \hline
Neurons per hidden layer & 64 \\ \hline
Discount factor & 0.995 \\ \hline
Entropy coefficient & $\delta$ = 0.01 \\ \hline
GAE parameter & $\lambda$ = 0.95\\ \hline
Recurrent sequence length & 64 \\ \hline
\end{tabular}
\caption{MARL learning parameters for the policy and value networks.}
\label{tab:RL_parameter}
\end{table}

\subsection{Performance analysis} 
Our analysis encompasses both the conflict graph abstraction and the more complex cellular network environments, providing insights into the effectiveness of our MARL-based approach across different network configurations. 

\subsubsection{QoS Performance in Conflict Graph}

\begin{figure} 
\centering
\includegraphics[width=\mywidth]{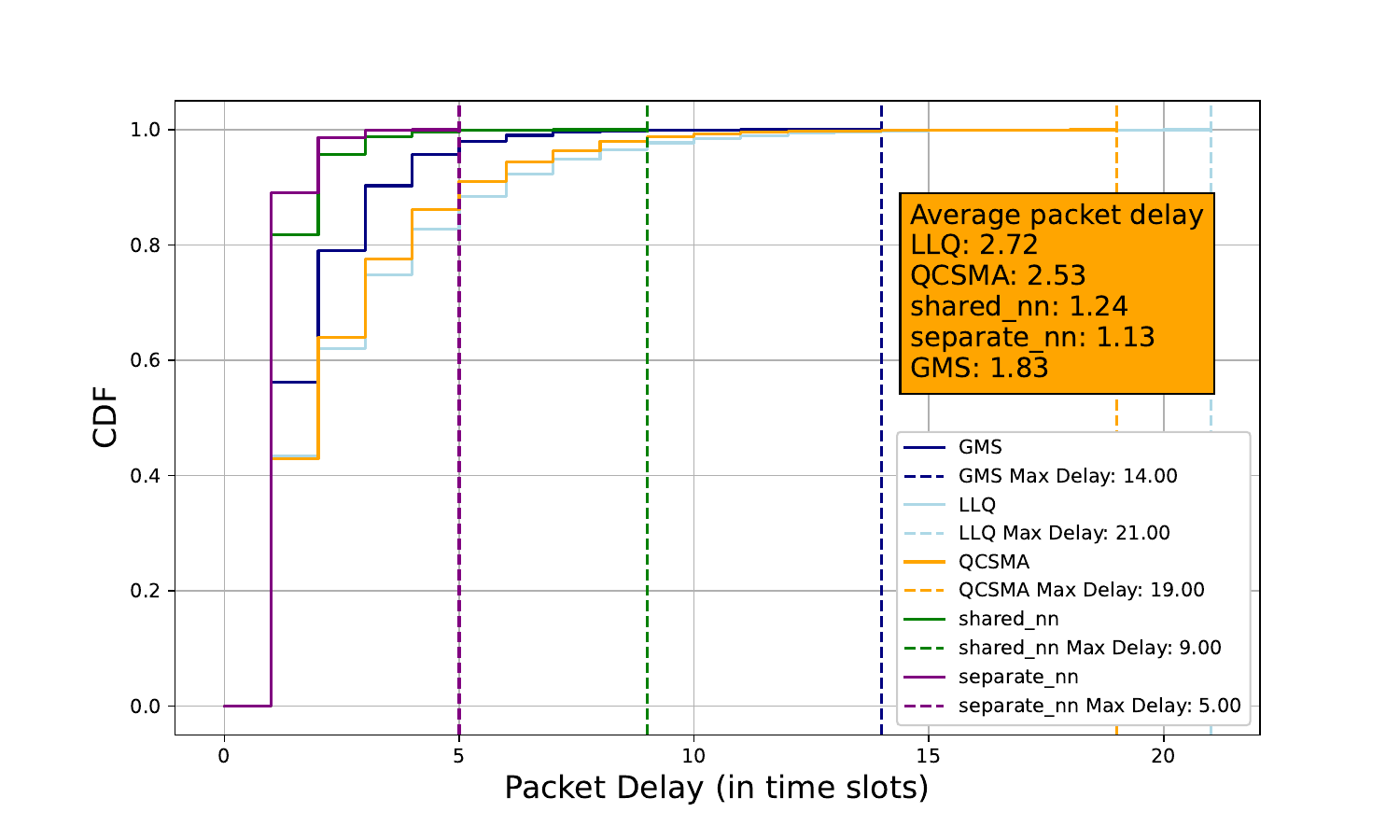}
\caption{CDFs of packet delays in the 8-device conflict graph.} 
\label{fig:8link_light_delay_cdf}
\end{figure}

\begin{figure} 
\centering
\includegraphics[width=\mywidth]{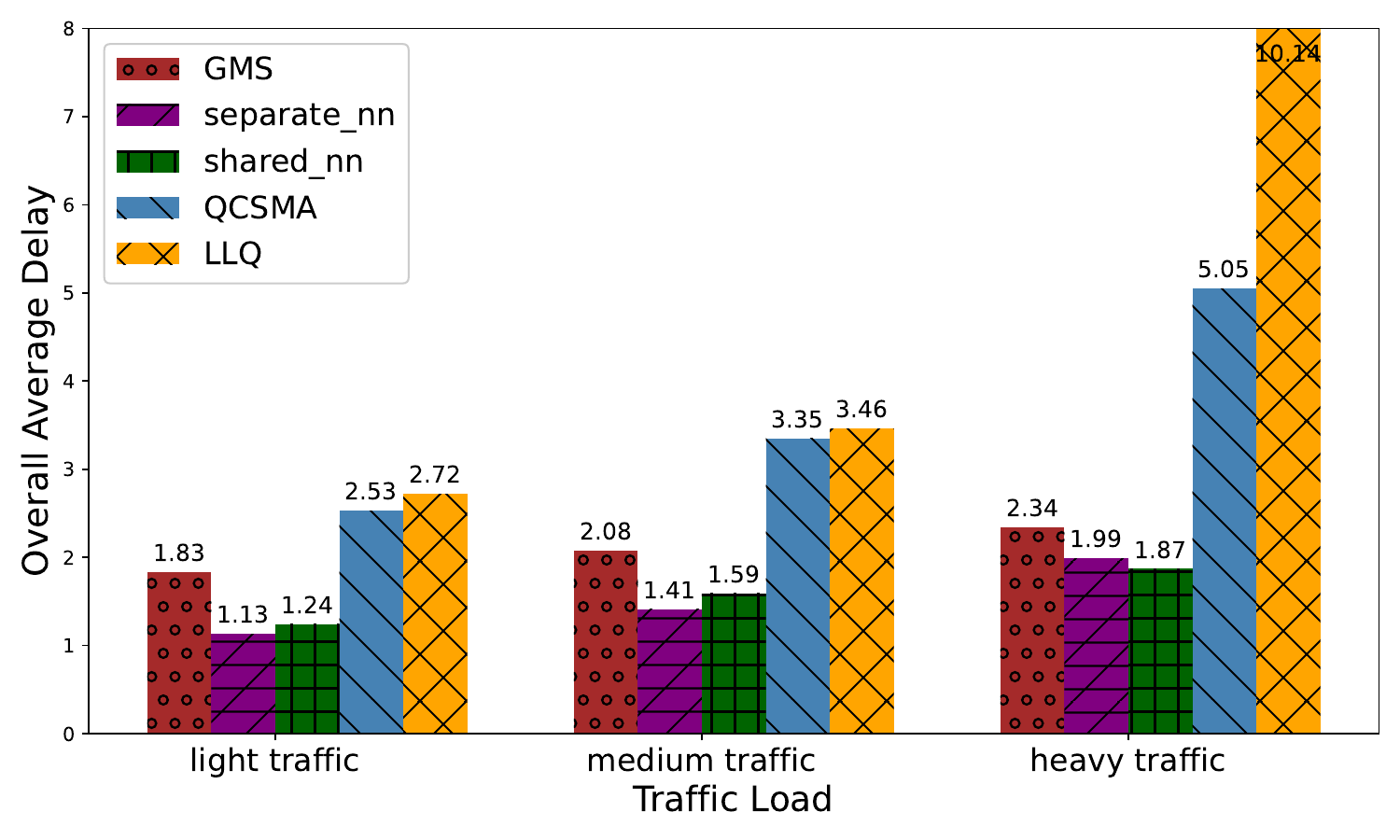}
\caption{The average packet delay of the 8-device conflict graph.}
\label{fig:8link_delay_cluster}
\end{figure}

We first examine the conflict graph scenario, a simplified abstraction of wireless network interactions. Here we compare the average packet delays achieved by our MARL method against the GMS, LLQ and Q-CSMA benchmarks. In this context, packet delay is measured in time slots, assuming successful transmission of one packet per time slot using one sub-band when scheduled conflict-free.

Fig.~\ref{fig:8link_light_delay_cdf} presents the cumulative distribution functions (CDFs) of packet delays under light traffic conditions in the conflict graph depicted by Fig.~\ref{fig:8 link conflict graph}. Both MARL approaches—utilizing shared or separate policies—outperform the benchmarks, with their CDFs dominating those of the other methods. Notably, over $80\%$ of packets are transmitted within a single time slot using either MARL method. 
The average packet delays of both MARL methods are lower than that of GMS and substantially lower than those of Q-CSMA and LLQ. Furthermore, MARL with separate policies achieves a significantly lower maximum packet delay compared to GMS, Q-CSMA, and LLQ.

We further test our algorithms under medium and heavy traffic conditions, which pose increased challenges to the learning method. Fig.~\ref{fig:8link_delay_cluster} illustrates the average packet delays across these scenarios. Under medium traffic conditions, the MARL-based solutions continue to outperform the benchmarks. In heavy traffic condition, which is relatively close to the boundary of the capacity region, the LLQ algorithm experiences high delays. The Q-CSMA scheduler's performance also degrades rapidly, while GMS remains stable and results in average packet delay of 2.34 time slots, whereas both MARL methods achieve average delays under 2 time slots.

The MARL methods demonstrate substantial improvements over benchmarks in terms of CDF, average delay, and maximum packet delay. A closer examination of the agents' policies reveals key differences: GMS and Q-CSMA transmit more conservatively, scheduling only devices in independent sets to avoid conflicts. In contrast, MARL methods operate in a richer action space, sometimes scheduling more aggressively than independent sets. Since conflicts occur between directional links, a subset of MARL-scheduled conflicting transmissions may still succeed.

\subsubsection{Accessible Information and Time Complexity}

\begin{table}
    \centering
    \begin{tabular}{|l|c|c|}
    \hline
    \textbf{Methods} & \textbf{Queue length} & \textbf{Broadcast}   \\  \hline
    GMS          & global & broadcast      \\ \hline
    LLQ           & local   & None          \\ \hline
    Q-CSMA   & local & broadcast     \\ \hline
    MARL shared      & local & None      \\ \hline
    MARL separate    & local & None  \\ \hline
    \end{tabular}
    \caption{Information needed for different methods in a conflict graph.}
    \label{tab:time complexity conflict}
\end{table}

Table~\ref{tab:time complexity conflict} summarizes the required information for different methods in the conflict graph setting. GMS is centralized and requires global queue length information. Q-CSMA necessitates some centralized coordination as each link sequentially sends a broadcasting signal to decide whether to participate in transmission. Table~\ref{tab:time complexity cellular} outlines the required information and execution times of the various methods in cellular network scenarios. The execution times are measured as the average time per execution over a testing episode in the network depicted by Fig.~\ref{fig:57link hexa deployment}) using a 4-core 2.8 GHz Core i7-1165G7 processor.

\begin{table}
\centering
\begin{tabular}{|l|r|r|r|}
\hline
\textbf{methods} & \textbf{queue lengths} & \textbf{CSI} & \textbf{execution time} \\ 
                 &                   &            & \textbf{$N=57, K=19, H= 3$} \\ \hline
FP               & global & global      & 58.13110 ms\\ \hline
WMMSE            & global & global      & 926.88220  ms \\ \hline
FITLinQ   & local & global     &  5.22326 ms\\ \hline
Greedy           & local &    None              &  0.16598 ms\\ \hline
MARL shared      & local & local      &  1.56926 ms\\ \hline
MARL separate    & local & local     &  1.30292 ms\\ \hline
\end{tabular}
\caption{Information exchange and execution time comparison for different methods in cellular network. In addition to queue lengths and CSI, FITLinQ also needs a broadcast signal similar to Q-CSMA.}
\label{tab:time complexity cellular}
\end{table}

\subsubsection{QoS Performance in Cellular Network}

\begin{figure} 
\centering
\includegraphics[width=\mywidth]{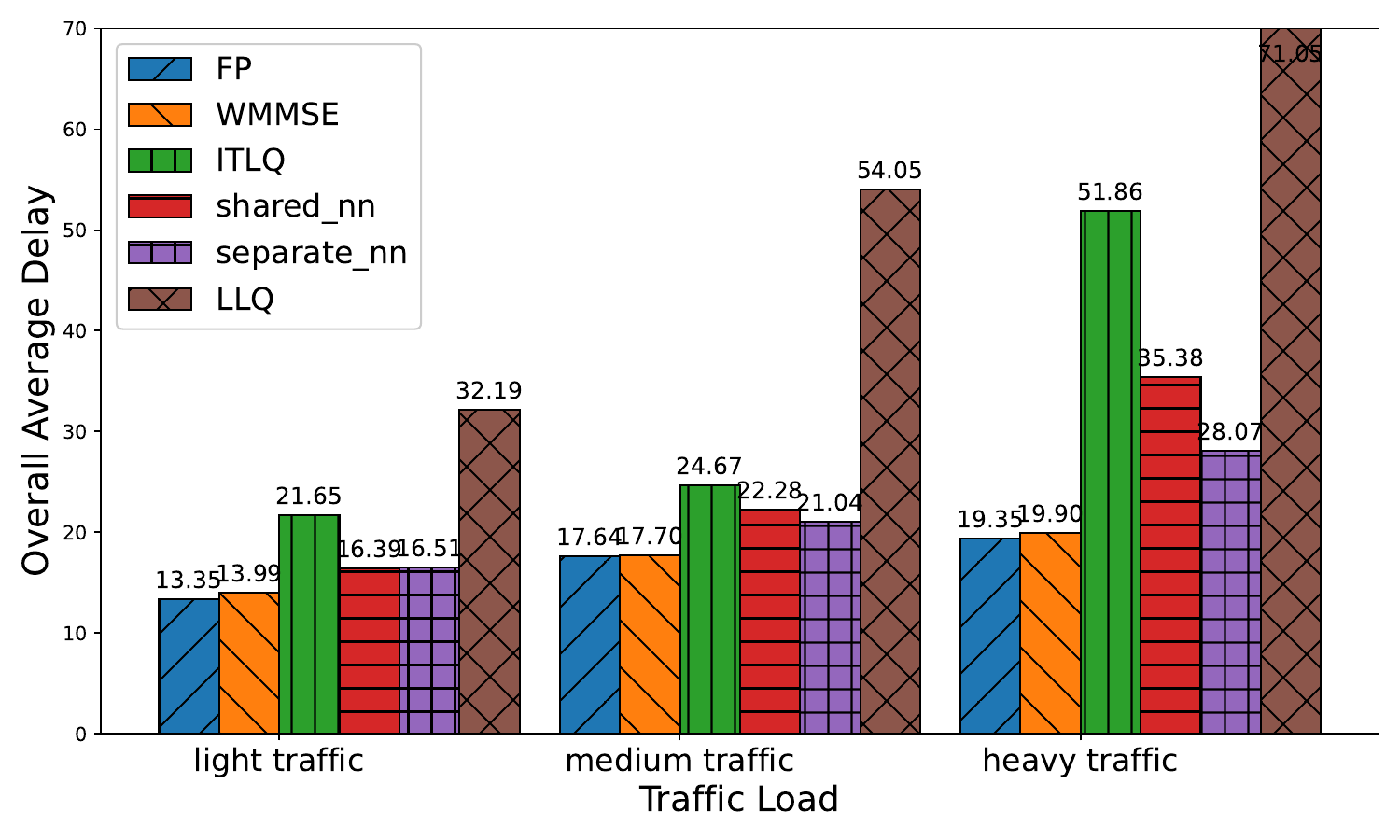}
\caption{QoS of the regular network depicted by Fig.~\ref{fig:57link hexa deployment}.}
\label{fig:57link_hexa_delay_cluster}
\end{figure}

\begin{figure} 
\centering
\includegraphics[width=\mywidth]{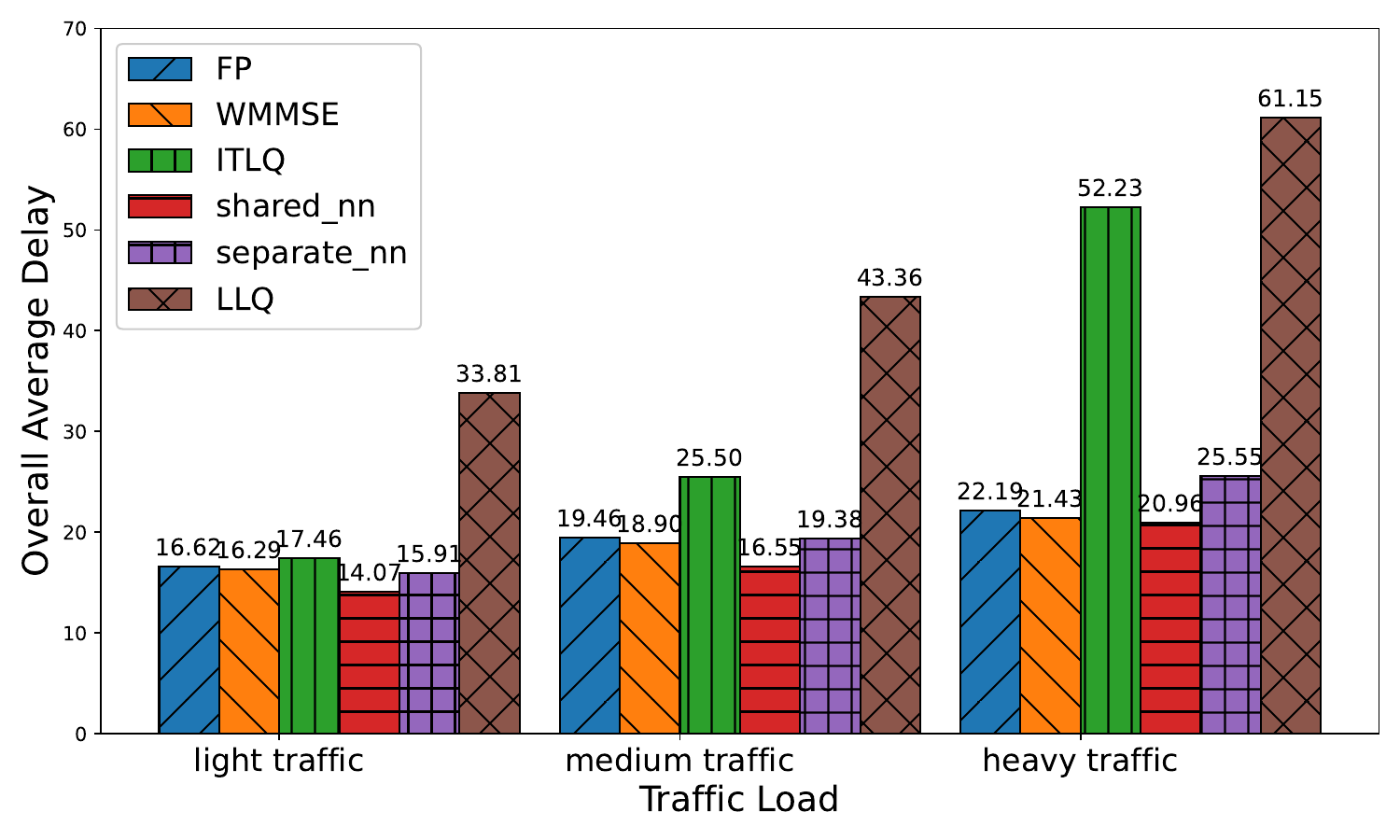}
\caption{QoS of the random network depicted by Fig.~\ref{fig:57link random deployment}.}
\label{fig:57link_random_delay_cluster}
\end{figure}

\begin{figure} 
\centering
\includegraphics[width=\mywidth]{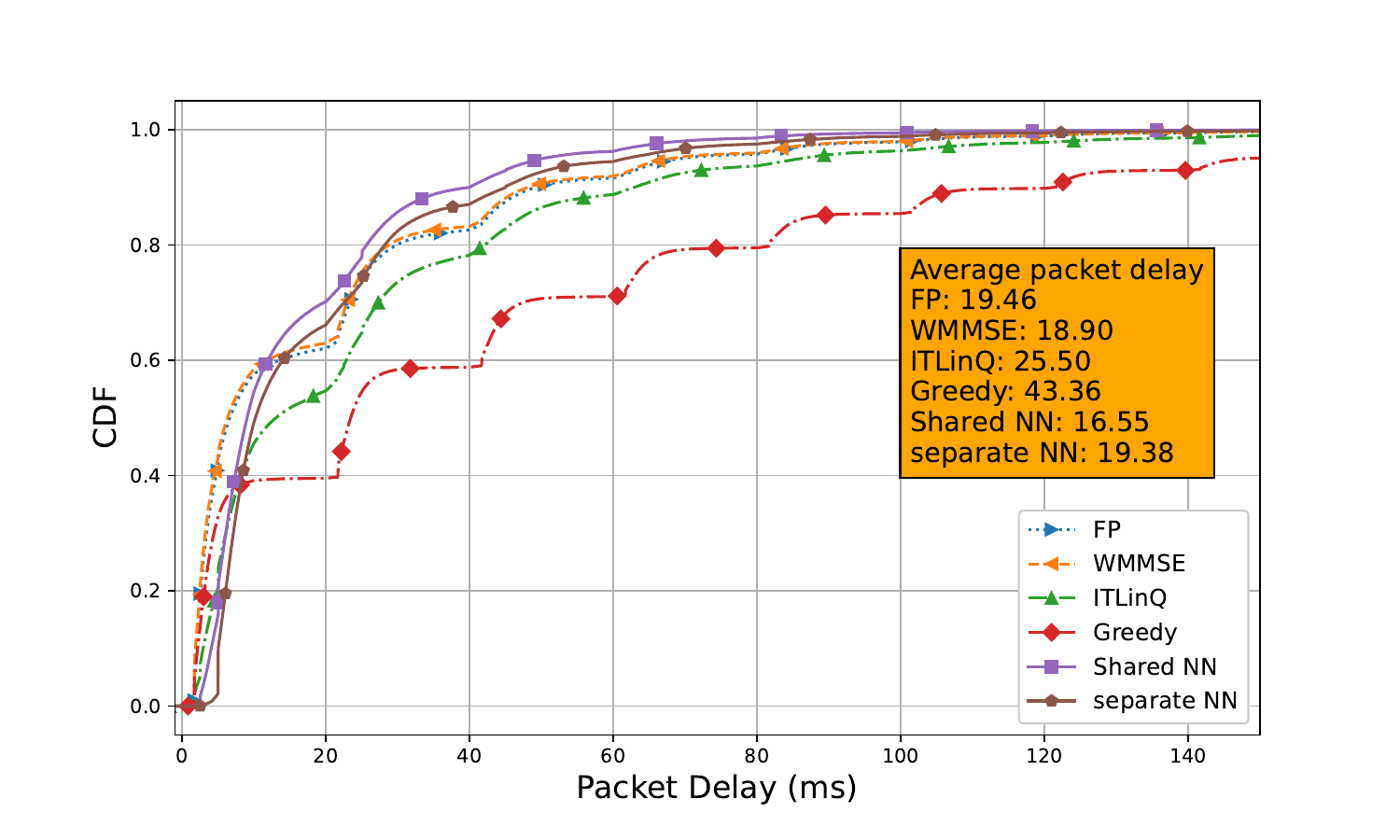}
\caption{CDFs of delays in the network depicted by Fig.~\ref{fig:57link random deployment}.} 
\label{fig:57link_medium_delay_cdf}
\end{figure}

We now evaluate our algorithm in more complex cellular network scenarios. Delays are measured in milliseconds, and the number of bits delivered in each transmission is determined by the SINR, generally not an integer number of packets. The packet delay is calculated once all of its bits are received. To validate the scalability, we test our MARL methods on relatively large networks consisting of 19 APs and 57 devices, as shown in Figs.~\ref{fig:57link hexa deployment} and~\ref{fig:57link random deployment}. As traffic intensity increases, benchmarks using local information degrade quickly, while our separate and shared policies remain stable and continue to outperform them, as illustrated in Figs.~\ref{fig:57link_hexa_delay_cluster} and~\ref{fig:57link_random_delay_cluster}. Compared to the genie-aided centralized methods, our methods offer similar performance in the network depicted by Fig.~\ref{fig:57link hexa deployment} and slightly better performance than FP and WMMSE in the network depicted by Fig.~\ref{fig:57link random deployment}. We plot the CDFs of packet delays for all successfully transmitted packets. As shown in Fig.~\ref{fig:57link_medium_delay_cdf}, the CDFs of our two methods clearly dominates the CDFs of the other 4 benchmarks. 

Our simulation results demonstrate that the proposed fully distributed MARL-based methods, using only local information, can achieve performance levels comparable to genie-aided centralized methods like FP and WMMSE. Notably, as traffic-driven approaches, our MARL-based solutions offer significant advantages in terms of real-time implementation. Table \ref{tab:time complexity cellular} shows that the execution time for our MARL methods is approximately 1-2 milliseconds, which is substantially smaller than the observed packet delays. This rapid execution enables real-time decision-making in dynamic network environments. In contrast, FP and WMMSE, being iterative optimization-based methods, require more and often unpredictable computational resources. Their execution times are one to two orders of magnitude larger than our MARL-based methods, making them challenging to deploy in real-time systems where rapid adaptation to changing network conditions is crucial.

Analysis of the agents' policies reveals that during training, they aim to balance utilizing as many sub-bands as possible for devices with long queues while avoiding excessive interference with neighbors based on local information.

\subsection{Policy Convergence} 

\begin{figure} 
\centering
\includegraphics[width=0.9\mywidth]{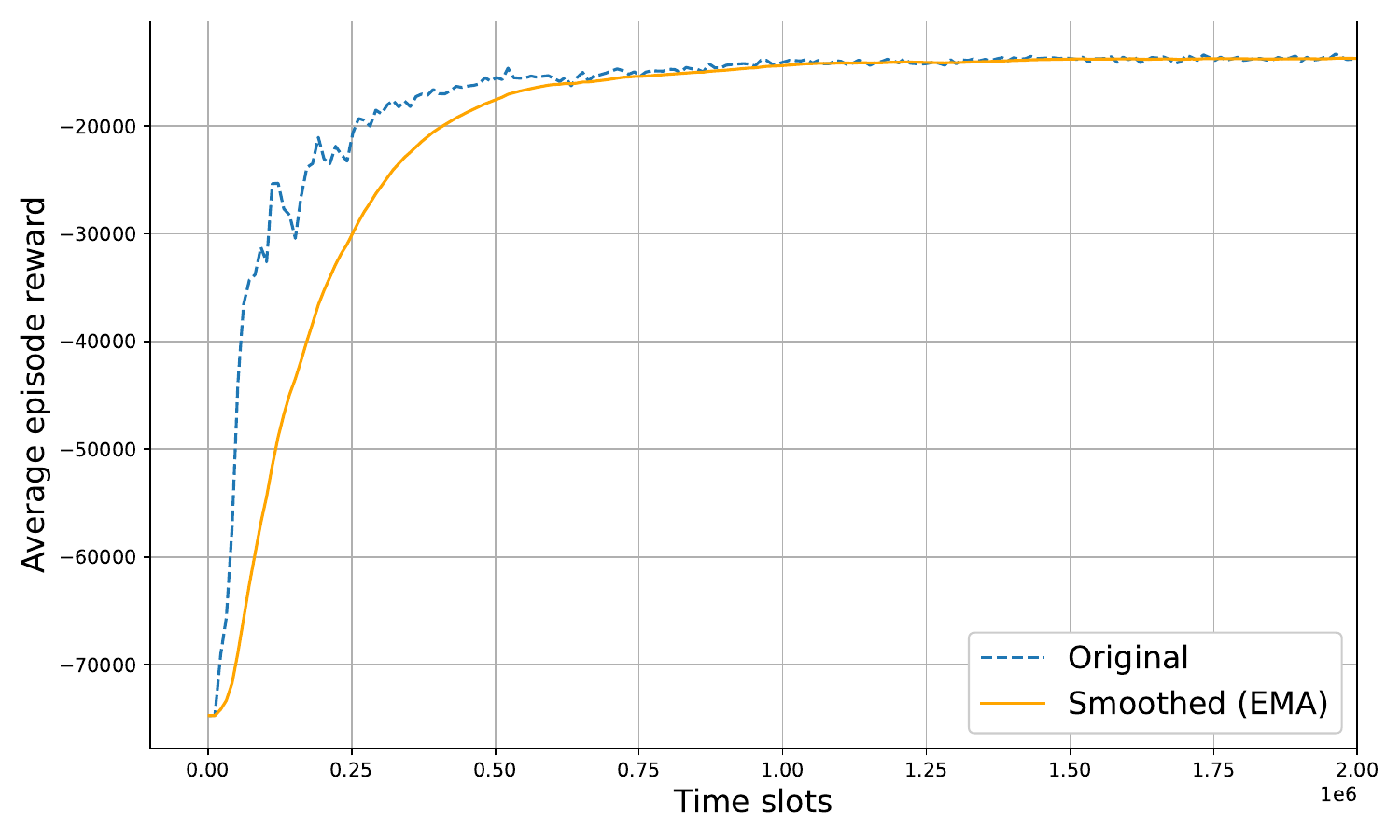}
\caption{Rewards of training episodes with shared policies.}
\label{fig:57link shared reward}
\end{figure}

\begin{figure}
\centering
\includegraphics[width=0.9\mywidth]{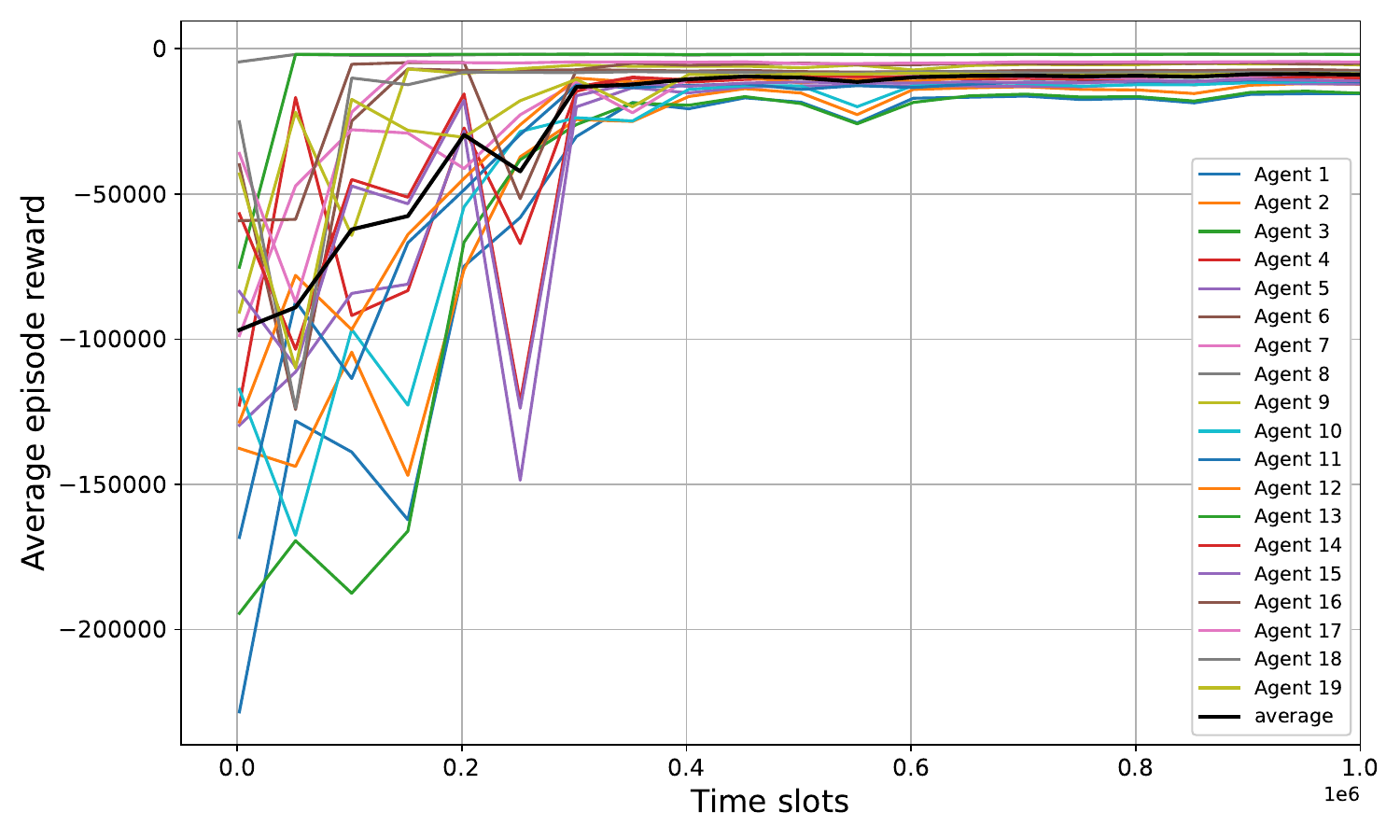}
\caption{Rewards of training episodes with separate policies.}
\label{fig:57link separate reward}
\end{figure}

To evaluate the training performance and algorithm convergence, we tested the learned policies every five training episodes and plotted the rewards. Fig.~\ref{fig:57link shared reward} illustrates the average reward from the MARL method using a shared policy for the 19-AP 57-device cellular network (Fig.~\ref{fig:57link hexa deployment}). The blue dashed line represents the average reward, with an exponential moving average curve in orange enhancing clarity. The reward initially improves rapidly, indicating quick learning by the agents. After approximately 750,000 time slots, the reward generally stabilizes, suggesting that each agent has successfully learned an effective and stable policy, resulting in a consistent and favorable cumulative reward.

Fig.~\ref{fig:57link separate reward} displays the rewards for the 19 agents using separate policies for the same network. During initial training, agents serving devices with low interference (e.g., agents 8 and 3, whose devices are not near cell boundaries) quickly achieve favorable rewards. Conversely, agents dealing with significant neighbor interference, like agent 5, face early challenges. Despite fluctuations, all agents' rewards generally trend upward, as evidenced by the average reward across all agents. They converge to efficient policies slightly faster than the shared policy approach, achieving convergence within approximately 600,000 time slots. These policies benefit individual agents and contribute to a stable and favorable cumulative reward for the entire network.

\subsection{Model Mismatch}

We examined the robustness of the MARL method when trained and tested under different traffic conditions. Performance is considered ``unstable'' if queue lengths persistently increase over time, ``good'' if it shows satisfactory QoS compared to the benchmark, and ``mixed'' if there is a combination of ``good'' and ``unstable'' results among the agents.
\begin{table} 
\centering
\begin{tabular}{|c|c|c|c|}
\hline
Traffic load for testing:
& Light & Medium & Heavy \\ \hline
if trained under light traffic                                     &         good       &         mixed       &         unstable       \\ \hline
if trained under medium traffic                                     &      good          &        good        &         unstable      \\ \hline
if trained under heavy traffic                             &     good           &      good          &        good        \\ 
\hline
\end{tabular}
\caption{Training and testing mismatch.}
\label{tab:mismatch}
\end{table}

Table~\ref{tab:mismatch} demonstrates that policies trained under heavier traffic loads exhibit better performance when handling lighter traffic loads. For instance, policies trained in heavy traffic loads demonstrate satisfactory behavior in both light and medium traffic environments. However, policies trained in light traffic show poor performance under medium and heavy traffic conditions.

\section{Conclusion}
\label{sec:Con}

We have introduced a novel traffic-driven MARL framework for resource allocation with QoS as the objective. We have proposed two MARL-based solutions: a fully distributed individual policy for each agent and shared policy for all agents. While the proposed solutions use only local information and require significantly less execution time,
numerical results demonstrate that we can achieve packet delay performance comparable to existing genie-aided centralized algorithms. The results also showcase the scalability and robustness of the trained policies across various network size and traffic conditions. The proposed framework is potentially applicable to a broader set of resource allocation problem.


\bibliographystyle{IEEEtran}
\bibliography{RLarxiv}

\begin{thebibliography}{10}
\providecommand{\url}[1]{#1}
\csname url@samestyle\endcsname
\providecommand{\newblock}{\relax}
\providecommand{\bibinfo}[2]{#2}
\providecommand{\BIBentrySTDinterwordspacing}{\spaceskip=0pt\relax}
\providecommand{\BIBentryALTinterwordstretchfactor}{4}
\providecommand{\BIBentryALTinterwordspacing}{\spaceskip=\fontdimen2\font plus
\BIBentryALTinterwordstretchfactor\fontdimen3\font minus \fontdimen4\font\relax}
\providecommand{\BIBforeignlanguage}[2]{{%
\expandafter\ifx\csname l@#1\endcsname\relax
\typeout{** WARNING: IEEEtran.bst: No hyphenation pattern has been}%
\typeout{** loaded for the language `#1'. Using the pattern for}%
\typeout{** the default language instead.}%
\else
\language=\csname l@#1\endcsname
\fi
#2}}
\providecommand{\BIBdecl}{\relax}
\BIBdecl

\bibitem{shi2011iteratively}
Q.~Shi, M.~Razaviyayn, Z.-Q. Luo, and C.~He, ``An iteratively weighted {MMSE} approach to distributed sum-utility maximization for a {MIMO} interfering broadcast channel,'' \emph{IEEE Transactions on Signal Processing}, vol.~59, no.~9, pp. 4331--4340, 2011.

\bibitem{shen2018fractional}
K.~Shen and W.~Yu, ``Fractional programming for communication systems—{Part I}: Power control and beamforming,'' \emph{IEEE Transactions on Signal Processing}, vol.~66, no.~10, pp. 2616--2630, 2018.

\bibitem{srikant2013communication}
R.~Srikant and L.~Ying, \emph{Communication Networks: An Optimization, Control, and Stochastic Networks Perspective}.\hskip 1em plus 0.5em minus 0.4em\relax Cambridge University Press, 2013.

\bibitem{tarjan1977finding}
R.~E. Tarjan and A.~E. Trojanowski, ``Finding a maximum independent set,'' \emph{SIAM Journal on Computing}, vol.~6, no.~3, pp. 537--546, 1977.

\bibitem{ni2011q}
J.~Ni, B.~Tan, and R.~Srikant, ``{Q-CSMA}: Queue-length-based {CSMA/CA} algorithms for achieving maximum throughput and low delay in wireless networks,'' \emph{IEEE/ACM Transactions on Networking}, vol.~20, no.~3, pp. 825--836, 2011.

\bibitem{naderializadeh2014itlinq}
N.~Naderializadeh and A.~S. Avestimehr, ``{ITLinQ}: A new approach for spectrum sharing in device-to-device communication systems,'' \emph{IEEE Journal on Selected Areas in Communications}, vol.~32, no.~6, pp. 1139--1151, 2014.

\bibitem{huang2006distributed}
J.~Huang, R.~A. Berry, and M.~L. Honig, ``Distributed interference compensation for wireless networks,'' \emph{IEEE Journal on Selected Areas in Communications}, vol.~24, no.~5, pp. 1074--1084, 2006.

\bibitem{kiani2007maximizing}
S.~G. Kiani, G.~E. Oien, and D.~Gesbert, ``Maximizing multicell capacity using distributed power allocation and scheduling,'' in \emph{2007 IEEE Wireless Communications and Networking Conference}.\hskip 1em plus 0.5em minus 0.4em\relax IEEE, 2007, pp. 1690--1694.

\bibitem{sun2018learning}
H.~Sun, X.~Chen, Q.~Shi, M.~Hong, X.~Fu, and N.~D. Sidiropoulos, ``Learning to optimize: Training deep neural networks for interference management,'' \emph{IEEE Transactions on Signal Processing}, vol.~66, no.~20, pp. 5438--5453, 2018.

\bibitem{zhao2021distributed}
Z.~Zhao, G.~Verma, C.~Rao, A.~Swami, and S.~Segarra, ``Distributed scheduling using graph neural networks,'' in \emph{Proc. IEEE International Conference on Acoustics, Speech and Signal Processing}.\hskip 1em plus 0.5em minus 0.4em\relax IEEE, 2021, pp. 4720--4724.

\bibitem{nasir2019multi}
Y.~S. Nasir and D.~Guo, ``Multi-agent deep reinforcement learning for dynamic power allocation in wireless networks,'' \emph{IEEE Journal on Selected Areas in Communications}, vol.~37, no.~10, pp. 2239--2250, 2019.

\bibitem{tan2020deep}
J.~Tan, Y.-C. Liang, L.~Zhang, and G.~Feng, ``Deep reinforcement learning for joint channel selection and power control in {D2D} networks,'' \emph{IEEE Transactions on Wireless Communications}, vol.~20, no.~2, pp. 1363--1378, 2020.

\bibitem{nasir2021deep}
Y.~S. Nasir and D.~Guo, ``Deep reinforcement learning for joint spectrum and power allocation in cellular networks,'' in \emph{2021 IEEE Globecom Workshops (GC Wkshps)}.\hskip 1em plus 0.5em minus 0.4em\relax IEEE, 2021, pp. 1--6.

\bibitem{khan2020centralized}
A.~A. Khan and R.~S. Adve, ``Centralized and distributed deep reinforcement learning methods for downlink sum-rate optimization,'' \emph{IEEE Transactions on Wireless Communications}, vol.~19, no.~12, pp. 8410--8426, 2020.

\bibitem{ge2023deep}
J.~Ge, Y.-C. Liang, L.~Zhang, R.~Long, and S.~Sun, ``Deep reinforcement learning for distributed dynamic coordinated beamforming in massive {MIMO} cellular networks,'' \emph{IEEE Transactions on Wireless Communications}, 2023.

\bibitem{chang2023federated}
H.-H. Chang, Y.~Song, T.~T. Doan, and L.~Liu, ``Federated multi-agent deep reinforcement learning (fed-madrl) for dynamic spectrum access,'' \emph{IEEE Transactions on Wireless Communications}, vol.~22, no.~8, pp. 5337--5348, 2023.

\bibitem{guo2020joint}
D.~Guo, L.~Tang, X.~Zhang, and Y.-C. Liang, ``Joint optimization of handover control and power allocation based on multi-agent deep reinforcement learning,'' \emph{IEEE Transactions on Vehicular Technology}, vol.~69, no.~11, pp. 13\,124--13\,138, 2020.

\bibitem{oliehoek2016concise}
F.~A. Oliehoek and C.~Amato, \emph{A Concise Introduction to Decentralized {POMDPs}}.\hskip 1em plus 0.5em minus 0.4em\relax Springer, 2016, vol.~1.

\bibitem{yu2022surprising}
C.~Yu, A.~Velu, E.~Vinitsky, J.~Gao, Y.~Wang, A.~Bayen, and Y.~Wu, ``The surprising effectiveness of {PPO} in cooperative multi-agent games,'' \emph{Advances in Neural Information Processing Systems}, vol.~35, pp. 24\,611--24\,624, 2022.

\bibitem{qu2020scalable}
G.~Qu, A.~Wierman, and N.~Li, ``Scalable reinforcement learning of localized policies for multi-agent networked systems,'' in \emph{Proceedings of the 2nd Conference on Learning for Dynamics and Control}, vol. 120.\hskip 1em plus 0.5em minus 0.4em\relax PMLR, 2020, pp. 256--266.

\bibitem{schulman2015high}
J.~Schulman, P.~Moritz, S.~Levine, M.~Jordan, and P.~Abbeel, ``High-dimensional continuous control using generalized advantage estimation,'' \emph{arXiv preprint arXiv:1506.02438}, 2015.

\end{thebibliography}

\end{document}